\documentclass{aa}  

\usepackage{graphicx}
\usepackage{txfonts}
\usepackage{lipsum}
\usepackage{subcaption}
\usepackage{color}
\usepackage{ulem}
\usepackage{amsmath}

\begin{document} 

    \title{An assessment of the Weak Field Approximation}

    \author{D. Arroyo Caballero\inst{1}\inst{2}\and 
    J. C. del Toro Iniesta\inst{1}\inst{2}\and
    D. Orozco Suárez\inst{1}\inst{2}\and  
    A. L. Siu-Tapia\inst{1}\inst{2}}
    \institute{Instituto de Astrofísica de Andalucía (CSIC), Apdo. de Correos 3004, E-18080 Granada, Spain \and
    Spanish Space Solar Physics Consortium (S$^3$PC), Spain}

    \date{\today}

    \abstract
        {Interpreting solar spectropolarimetric observations requires understanding of how variations in atmospheric parameters affect the formation of spectral lines. The Weak Field Approximation (WFA) provides a simple and widely used method to infer magnetic fields from Stokes profiles, but its range of validity and sensitivity to atmospheric conditions are not always well quantified.}
        {We study the reliability and limitations of the WFA using Generalized Response Functions (GRFs), which allow us to quantify how the magnetic field inferred by the WFA responds to perturbations in different atmospheric parameters.}
        {We derive analytical expressions for the GRFs of the WFA and compute them from synthetic Stokes profiles generated with the SIR and DeSIRe radiative transfer codes. The analysis is performed for a set of photospheric and chromospheric lines under different magnetic field strengths and inclinations. By studying these GRFs, we evaluate the sensitivity of the WFA to the atmospheric structure and assess its effective range of validity.}
        {The GRFs reveal that the magnetic field inferred through the WFA depends on the atmospheric structure, the magnetic-field inclination, and the spectral region used in the analysis. For the photospheric lines considered here, the approximation progressively departs from the actual magnetic field as the field strength increases, well before reaching the saturation field, as theoretically expected. This behavior is related to the loss of linearity between the Stokes $V$, $Q$ and $U$ profiles and the derivatives of the intensity profile. The results also show that the transverse-field inference is particularly sensitive to the wavelength range (as it was originally formulated), with the line wings providing more reliable estimates. For the chromospheric lines considered here, the line core and wings clearly probe different atmospheric layers, with the wings providing reliable estimates of the photospheric magnetic field.}
        {Our analysis shows that the effective range of validity of the WFA is more limited than commonly assumed. The approximation begins to depart from the actual magnetic field well before reaching the so-called saturation value, and its reliability depends on the magnetic-field inclination and on the wavelength range used for the inference. In particular, the choice of wavelength interval is critical for the transverse component, while the wings of chromospheric lines can provide useful information on the photospheric magnetic field. These results are consistent with the original perturbative formulation of the WFA.}
   

    \maketitle

\section{Introduction} \label{sec:introduction}

The Weak Field Approximation (WFA) is a simple and widely used method to infer magnetic fields from spectropolarimetric observations. To our knowledge, the fundamental idea was first introduced by \cite{Semel 1967}, where the circular polarization signal was shown to be proportional to the derivative of the intensity profile under the assumption that the Zeeman splitting is much smaller than the Doppler width of the spectral line. Later, \citet{Landi 73} provided a rigorous mathematical formulation of the approximation through a perturbative treatment of the radiative transfer equation (RTE), establishing the relations between the Stokes parameters and the magnetic field that are still commonly used today.\footnote{According to \citet{Landi 73}, their results generalize those from other authors who partially obtained similar results earlier even than \cite{Semel 1967}. Unfortunately, those papers are written in Russian and out of reach to the authors of the present paper.}

Because of its analytical simplicity and extremely low computational cost, the WFA has been widely used in solar spectropolarimetry. In particular, it provides a direct estimate of the longitudinal and transverse components of the magnetic field from the observed Stokes profiles, avoiding the computationally expensive inversion techniques required by more sophisticated diagnostics. Historically, these advantages made the WFA the basis of the so-called magnetograph formula, which was frequently employed when only one or two wavelength samples of Stokes $I$ and $V$ were available and more advanced radiative-transfer-based techniques had not yet been developed. Despite the widespread availability of modern spectropolarimetric instruments and efficient radiative transfer inversion codes, the WFA remains extensively used. It has been successfully applied to the analysis of magnetic fields in both the quiet Sun (e.g., \citealt{Martinez Gonzalez 09}) and active regions (e.g., \citealt{Kleint 17, Ferrente 23}). Its simplicity makes it particularly attractive for the analysis of large datasets and high-cadence observations, for which inversion-based approaches may still be prohibitively time-consuming. Furthermore, when combined with other simple diagnostics such as the center of gravity method, the WFA has been recommended \citep{del Toro 16} as an objective initialization for inversion codes, significantly improving their convergence properties \citep{Cobos 2015}.

Although the WFA is mostly used for photospheric lines, it has also been employed for interpreting chromospheric lines \citep[e.g.,][]{Pillet 90, Siu Tapia 25a, Siu Tapia 25b}. Its applicability for some of these lines, which are formed under non-local thermodynamic equilibrium conditions, has also been tested \citep{Alfonso 23}. Their typically broad line profiles make them especially good candidates for the approximation. For instance, \cite{Centeno 18} studied the validity and limitations of the WFA for the \ion{Ca}{ii}~8542~\AA\/ chromospheric line. In chromospheric applications, it is common to treat separately the spectral line core and wings, since they are thought to sample different atmospheric layers and may satisfy the weak-field conditions differently (e.g., \citealt{Centeno 18, Siu Tapia 20}). Specifically, they demonstrated that the validity of this approximation depends on the magnetic field component to be inferred, whether longitudinal or transverse. They also showed that the line-of-sight component of the magnetic field vector can be retrieved from the core of the spectral line alone and is reliable for field strengths up to 1200 G. However, the transverse component of the magnetic field is significantly more challenging to infer using the WFA. Due to the weak linear polarization signals and the non-LTE formation of the \ion{Ca}{ii}~8542~\AA\/ line, the approximation becomes less reliable for retrieving this component, especially in regions with weak or complex magnetic fields.

The WFA relies on several assumptions that limit its applicability. See, e.g., \cite{Shchukina 19} for a careful study of the validity in simulated cases of truly weak fields.  In particular, the approximation assumes that the vector magnetic field and the line-of-sight (LOS) velocity are constant throughout the atmosphere within the region where the spectral line is formed. In realistic solar atmospheres such as sunspot penumbrae, this condition is rarely satisfied, since both the magnetic field and the LOS velocity generally vary with height and may present significant gradients. Moreover, large variations of the magnetic field with height are generally found in weak-field regions, according to radiative MHD simulations. There, convective motions continuously drag, stretch, and twist the magnetic field lines. In contrast, strong-field regions, such as sunspot umbrae, suppress convective motions, leading to a comparatively smaller variations with height \citep{Stein 12}. 

On top of that, the actual range of validity of the WFA is often overestimated. As we show in the following sections, the approximation breaks down earlier than is commonly assumed, even in regimes where it is frequently applied. This limitation becomes even more critical when estimating the transverse component of the magnetic field, since it arises from higher-order terms in the RTE expansion in terms of the ratio between the Zeeman splitting and the Doppler width of the line, and is therefore intrinsically more sensitive to deviations from the weak-field regime.

Despite its widespread use, a systematic assessment of these limitations across different spectral lines and atmospheric conditions is still lacking. In this work we revisit the validity of the WFA and explore its limitations in detail. In particular, we investigate how the inferred magnetic field depends on the actual atmospheric parameters and on the properties of the spectral lines used. No artificial noise is added to the synthetic Stokes profiles and no convolution with the instrumental profiles is applied either. This is intended to highlight results---strengths and limitations---that strictly depend on radiative transfer. Specifically, no analysis is made for the dependence of the results on photon noise.

\section{Customary formulation of the WFA} \label{sec:formulation}

Within the weak field regime, the monochromatic Stokes parameters of a given spectral line of central wavelength $\lambda_0$ satisfy analytic relations with the magnetic field components, provided that some assumptions hold. Following \citet{Landi 04}, the theoretical expressions for the practical application of the so-called Weak Field Approximation are:
\begin{equation}
    \label{eq:WFA_formal}
    \begin{aligned}
        V(\lambda) & = -\Delta\lambda_B\,\overline{g}\,\cos\gamma\,\frac{\partial I(\lambda)}{\partial \lambda}, \\[6pt]
        Q(\lambda_0) & = -\frac{1}{4}\,\Delta\lambda_B^2\,\overline{G}\,\sin^2\gamma\, \left. \frac{\partial^2 I(\lambda)}{\partial \lambda^2} \right|_{\lambda=\lambda_0},
    \end{aligned}
\end{equation}
where $I(\lambda)$, $Q(\lambda)$, and $V(\lambda)$ are Stokes parameters, $\gamma$ is the magnetic-field inclination with respect to the LOS, $\overline{g}$ is the effective Landé factor of the line, $\overline{G}$ is another parameter easily calculated from the quantum numbers of the line, and $\Delta\lambda_B = 4.67 \cdot  10^{-13} \lambda_0^2 B$ is the Zeeman splitting for a Landé factor of unity. These expressions assume a uniform field of strength $B$---expressed in G---and azimuth angle $\varphi = 0^\circ$; $\lambda_0$ is expressed in \AA.\footnote{If $\varphi \neq 0$, then $Q$ should be substituted by $L \equiv \sqrt{Q^2 + U^2}$.} 

\begin{figure*}
    \centering
    \includegraphics[width=\hsize]{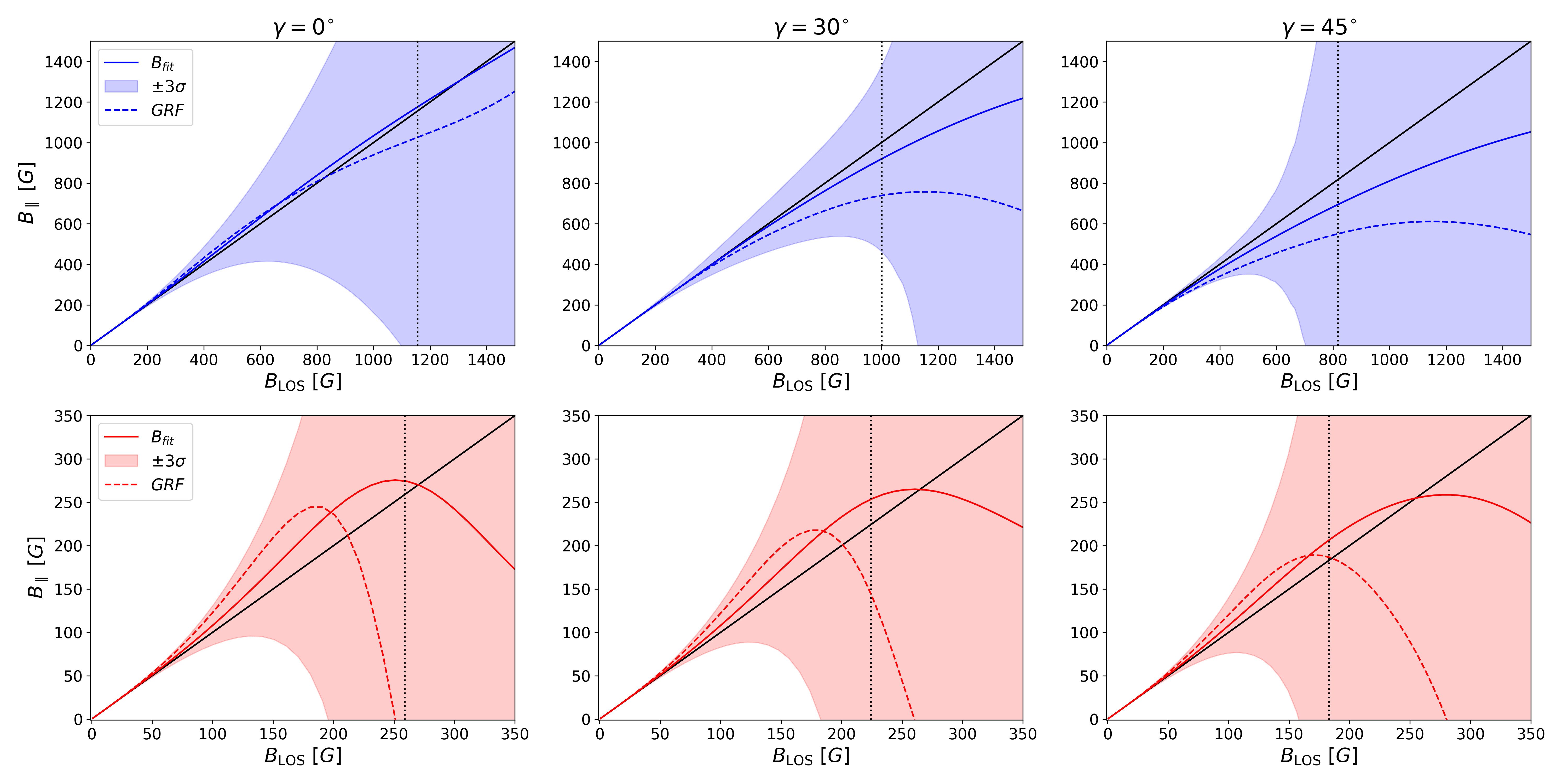}
    \caption{Inferred $B_{\parallel}$ magnetic field component vs.\ model $B_\mathrm{LOS}$ (solid, colored lines) for three inclinations, obtained through Eq.~(\ref{eq:B_WFA_P}) for \ion{Fe}{i} 6301.5~\AA~ (blue) and 15648~\AA~ (red), including fitting uncertainties (shadowed regions). The solid black line represents the bisector of the diagram. The dotted vertical lines indicate $B_{\rm sat} \cos{\gamma}$ from Eq.~(\ref{eq:B_sat}). Dashed curves show the expected results from the integrated GRFs (see Section~\ref{sec:longitudinal}).}
    \label{fig1}
\end{figure*}

Equations~(\ref{eq:WFA_formal}) can be rewritten in terms of the WFA longitudinal and transverse magnetic field components, $B_{\parallel}$ and $B_{\perp}$:
\begin{equation}
    \label{eq:WFA_B}
    \begin{aligned}
        V(\lambda) & = -C\,\overline{g}\,B_{\parallel}\,\frac{\partial I(\lambda)}{\partial \lambda}, \\[6pt]
        Q(\lambda_0) & = -\frac{1}{4}\,C^2\,\overline{G}\,B_{\perp}^2\, \left. \frac{\partial^2 I(\lambda)}{\partial \lambda^2} \right|_{\lambda=\lambda_0}.
    \end{aligned}
\end{equation}
where $C \equiv \Delta\lambda_B / B$. From here on, we shall distinguish between $B_\parallel$ and $B_\perp$ for the retrieved estimates and the true $B_\mathrm{LOS}$ and $B_\mathrm{T}$.

In practice, since the expression for the linear polarization is only valid at the core of the line, one has to rely on further assumptions \citep{Landi 04} to find an expression useful in the wings:
\begin{equation}
    \label{eq:WFA_Q_wings}
    Q(\lambda) \approx \frac{3}{4}\,C^2\,\overline{G}\,B_{\perp}^2\,\frac{1}{\lambda-\lambda_0}\,\frac{\partial I(\lambda)}{\partial \lambda},
\end{equation}
This approximate relation fails for $\lambda = \lambda_0$, i.e., around the line core, where only the general expression with the second derivative should be used. 

While the previous expressions provide the theoretical framework to estimate the magnetic field at single wavelengths, in real observations the Stokes parameters are sampled at $n$ wavelengths across the spectral line. Even in the ideal case that Eqs.~(\ref{eq:WFA_B}) and (\ref{eq:WFA_Q_wings}) would perfectly hold for all the wavelength samples across the line profile to provide the same $B_\parallel$ and $B_\perp$, which---we advance---it is not the case, noise would introduce variations among samples. To obtain more robust estimations, the (constant) magnetic field is inferred, for instance, by performing a least-squares fit using all available (or useful) wavelength samples \citep{Martinez Gonzalez 09}:
\begin{equation}
    \label{eq:B_WFA_P}
    B_{\parallel} = -\frac{1}{C\,\overline{g}}\,\frac{\sum_{i=1}^n V_i\,I_i^\prime}{\sum_{i=1}^n (I_i^\prime)^2}, 
\end{equation}

\begin{equation}
    \label{eq:B_WFA_T}
    \begin{aligned}
    B_{\perp}^2 \left(\lambda_{\mathrm{core}} \right) & = -\frac{4}{\,C^2\,\overline{G}}\,\frac{\sum_{i=1}^{n_{\rm core}} Q_i\,I_i^{\prime\prime}}{\sum_{i=1}^{n_{\rm core}} \left(I_i^{\prime\prime}\right)^2}, \\[6pt]
    B_{\perp}^2 \left(\lambda_{\mathrm{wing}} \right) & = \frac{4}{3\,C^2\,\overline{G}}\,\frac{\sum_{i=1}^{n_{\rm wing}} Q_i\,\frac{I_i^\prime}{(\lambda_i-\lambda_0)}}{\sum_{i=1}^{n_{\rm wing}} \left(\frac{I_i^\prime}{\lambda_i-\lambda_0}\right)^2},
    \end{aligned}
\end{equation}
where $i$ runs for the sampled wavelengths and the prime denotes partial differentiation with respect to $\lambda$. Certainly, an operational definition of $n_{\rm core}$ and $n_{\rm wing}$ is needed for the practical application of formulae (\ref{eq:B_WFA_T}), although ideally $n_{\rm core}$ should be limited to $\lambda = \lambda_0$. 

The WFA indeed requires the Zeeman splitting to be much smaller than the Doppler width of the line: $\overline{g}\Delta\lambda_B \ll \Delta\lambda_D$ or $\overline{g} v_B \ll 1$, where $v_B \equiv \Delta\lambda_B/\Delta\lambda_D \ll 1$. This condition allows one to derive an expression of a saturation field $B_{\rm sat}$ (\citealt{Landi 04}) to estimate the breakdown point:
\begin{equation}
    \label{eq:B_sat}
    B_{\mathrm{sat}} \equiv \frac{\sqrt{1.663\times10^{-2}\,\frac{T}{\mu} + \xi_0^2}}{1.4\times10^{-7}\,\overline{g}\lambda_0},
\end{equation}
where $T$ is temperature, $\mu$ the atomic weight, and $\xi_0$ the microturbulence. The temperature is taken at $\log\tau = 0$, although it would be more accurate to use the (smaller) temperature at which the line is most sensitive to the magnetic field. Nevertheless, it is just an illustrative value of the upper limit where the WFA is no longer valid. That is, $B \ll B_{\rm sat}$ is necessary (not sufficient) for the WFA to hold, according to \cite{Landi 04}. In practice, some authors have used $B < B_{\rm sat}$ without further justification \citep{Asensio Ramos 11} or assumed $\overline{g} \Delta\lambda_B/\Delta\lambda_D = 0.5$ as a good upper limit \citep{Afonso 25}. We will test these limits quantitatively below. Naturally, $B_{\mathrm{sat}}$ is inversely proportional to $\overline{g}\lambda_0$, that is, to the Zeeman sensitivity and the central wavelength of the line. 

Noteworthy is the fact that, just as Eqs.~(\ref{eq:WFA_formal}), (\ref{eq:WFA_B}) and (\ref{eq:WFA_Q_wings}) are monochromatic and therefore strictly valid at a single wavelength, Eq.~(\ref{eq:B_sat}) only holds at a single optical depth, namely that corresponding to temperature $T$. Academically, this is enough (the approximation is mathematically correct) under the assumption of $\vec{B} =$ constant. However, if one is using the WFA to retrieve gradients between the photosphere and the chromosphere by separating the core and the wings of the lines, at least two different temperatures should be used, as $\Delta\lambda_D$ changes through the atmosphere, and therefore two $B_\mathrm{sat}$ values must be estimated, one for the narrower core and another for the broader wings. This is typically overlooked by observers.

To illustrate the performance of the WFA, we compute synthetic Stokes profiles for two Zeeman-sensitive photospheric lines (\ion{Fe}{i} 6301.5~\AA\ and \ion{Fe}{i} 15648~\AA, with $\overline{g} = 1.67$ and 3, respectively), using the synthesis mode of SIR \citep{Ruiz Cobo 92} with the HSRA model atmosphere \citep{Gingerich 71} with added constant magnetic field strength, inclination, and azimuth, and zero LOS velocity. We vary the strength and inclination, and apply Eq.~(\ref{eq:B_WFA_P}) to retrieve $B_{\parallel}$. 

Figure~\ref{fig1} displays the inferred $B_{\parallel}$ (with fit uncertainties) versus the true $B_\mathrm{LOS}$ for three inclinations $(\gamma = 0^\circ, 30^\circ, 45^\circ)$. The inferred values are represented in blue or red solid lines, to be compared with the bisectors in solid black lines. 

At $\gamma=0^\circ$ (purely longitudinal field), the approximation appears to reproduce quite well the model longitudinal component up to values close to, or even larger than, $B_{\rm sat}$, marked by the dotted vertical lines. The IR line deviates more as a consequence of having a comparatively larger Zeeman splitting than the visible one. The behavior of our example lines is consistent with that found by \cite{Centeno 18} for the \ion{Ca}{ii} line at 8542 \AA, with a smaller Zeeman sensitivity ($\overline{g} = 1.1$). Both Centeno's results and ours could be said, thus, to be in contradiction with the theoretical development of the WFA, which requires $B \ll B_{\mathrm{sat}}$. 

A glance at the middle and right panels of Fig.~\ref{fig1}, clearly reveals that, in fact, the theoretical requirement makes sense: the agreement between the numerical estimates and the true longitudinal fields breaks at smaller $B_\mathrm{LOS}$. A neat dependence of our exercise on $\gamma$ is displayed. Indeed, the saturation value in the plot decreases as $B_\mathrm{sat} \cos{\gamma}$. Such a dependence was not explored by Centeno or, to the best of our knowledge, by anybody else, and indeed encloses the reason for the seeming contradiction, as we shall see later (see Section~\ref{sec:revisited}).

The uncertainties of the numerical calculations grow exponentially as reflected by the shaded areas. These uncertainties arise solely from the fit, and are independent of any instrumental noise. The origin of these large uncertainties can be traced to the breakdown of the fundamental assumption of the WFA, namely the linear relationship between Stokes $V$ and the derivative of the intensity profile. As discussed by \citet{del Toro 16}, the modification of the Stokes $I$ profile by the vector magnetic field introduces departures from this linear behavior (see also Section~\ref{sec:revisited}). To investigate whether the increasing uncertainties seen in Fig.~\ref{fig1} are related to such a loss of linearity, we examine the relation between $V$ and $I'$ in the synthetic profiles.

\begin{figure*}
    \centering
    \includegraphics[width=\hsize]{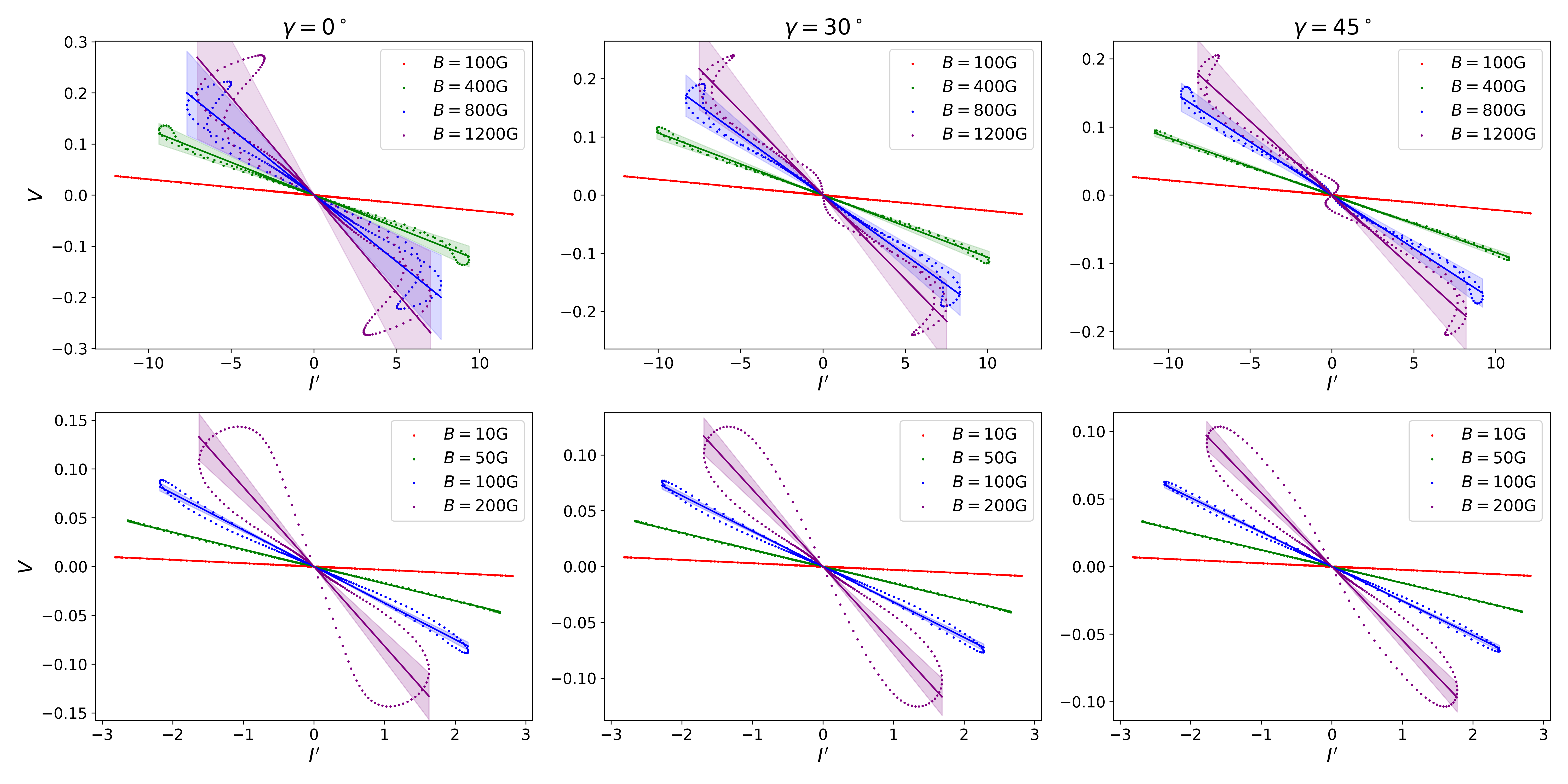}
    \caption{Scatter of $V$ vs.\ $I'$ for \ion{Fe}{i} 6301.5~\AA~ and 15648~\AA~ at various $B$'s and inclinations. Solid lines are least-squares fits (Equation~\ref{eq:B_WFA_P}).  The loss of linearity with increasing $B_\mathrm{LOS}$ is evident.}
    \label{fig2}
\end{figure*}

In Fig.~\ref{fig2}, we show a scatterplot of $V$ versus $I'$ for both lines and different fields and inclinations. At low fields, the points lie on a straight line, confirming $V \propto I'$ (as expected in the WFA). As $B$ increases, the relationship becomes highly nonlinear. An 8 shape or inverted-s 8 shape starts to develop, which is really apparent at the saturation field strength. In particular, near the inflection point of $I(\lambda)$ (where $|I'|$ is largest), the data scatter away from the straight line. This nonlinearity generates the large error areas in Figure~\ref{fig1}. If only those points near the inflection had been taken, the WFA would produce polarity changes! In fact, using different wavelength samples will return completely different $B_\parallel$. When the inclination is 30$^{\circ}$ or 45$^{\circ}$, clear nonlinearities also appear at the core of the line (where $I^\prime \simeq 0$). Thus, even when the fit slope may accidentally match the true longitudinal field for some cases, the result has no physical meaning since the linear model no longer applies beyond the weak regime, and moreover depends on the field inclination. Note that these are the easiest cases one can conceive, as the model atmosphere has constant vector magnetic field and LOS velocities. In real applications, where both $\Vec{B}$ and $v_{\rm LOS}$ significantly vary with depth, departures from linearity can reasonably be expected much larger than those in our numerical example, and moreover, they are not known in advance and can lead to misleading results. Indeed, the specific shape of the $V$ vs. $I'$ curve was already detected quite some time ago: \cite{Pillet 90} found that shape in \ion{Ca}{ii} K line observations. 

The reason for this behavior lies on the dependence of Stokes $I$ on $\Vec{B}$, which is often obliterated, as we shall understand in Section~\ref{sec:revisited}.

\section{Response Functions as a diagnostic tool} \label{sec:results}

Response functions, hereafter referred to as RFs, quantify the sensitivity of the Stokes profiles to small perturbations in the atmospheric quantities, such as temperature, magnetic field vector, or LOS velocity. They were first introduced in the literature for polarized light by \cite{Landi 77}. Their analytic expression in terms of the formal solution of the radiative transfer equation was presented by \citet[][see also \citeauthor{Ruiz Cobo 94} \citeyear{Ruiz Cobo 94}]{Almeida 92}.

Generalized response functions (GRFs), introduced by \citet{Ruiz Cobo 94}, extend this concept to any observed parameter that can be linearly derived from the Stokes profiles. Since RFs act as partial derivatives of the Stokes profiles with respect to the atmospheric model parameters (solar physical quantities at given optical depths, $x_j$),\footnote{Index $j$ runs over all the physical quantities relevant to line formation multiplied by the number of grid points in the discretized atmosphere.} applying the chain rule, GRFs allow us to quantify the sensitivity of such observed parameters to perturbations in the atmospheric ones.

Since atmospheric quantities are stratified in the solar atmosphere, the model parameters $x_j$ are defined at specific optical depths. Consequently, RFs are intrinsically depth-dependent, describing how a perturbation of a given physical quantity at a particular optical depth $\tau$ modifies the emergent Stokes profiles. In this sense, RFs provide a measure of the vertical sensitivity of the spectral line formation process, and any quantity derived from them, including the GRFs, naturally inherits this dependence on optical depth.

As shown in Sect.~2, the longitudinal or transverse magnetic field inferred using the WFA are among these observed parameters, and therefore its GRFs can be computed. Differentiating the longitudinal field $B_{\parallel}$ from Eq.~(\ref{eq:B_WFA_P}) with respect to a parameter $x_j$, its generalized response to perturbations in $x_j$ reads
\begin{equation}
    \label{eq:R_BL}
    \begin{split}
    R^{B_{\parallel}}_{x_j} = \frac{\partial B_{\parallel}}{\partial x_j} = 
    -\frac{1}{C\,\overline{g}}\,\left[\frac{\sum_{i=1}^n \left( \frac{\partial V_i}{\partial x_j}\,I_i' + V_i\,\frac{\partial I'_i}{\partial x_j} \right)}{\sum_{i=1}^n (I'_i)^2} -\right. \\ \left. \frac{2\sum_{i=1}^n V_i I'_i \sum_{i=1}^n I'_i \frac{\partial I'_i}{\partial x_j}}{\left[\sum_{i=1}^n (I'_i)^2\right]^2}\right].
    \end{split}
\end{equation}

\begin{figure}[ht]
    \centering
    \includegraphics[width=\hsize]{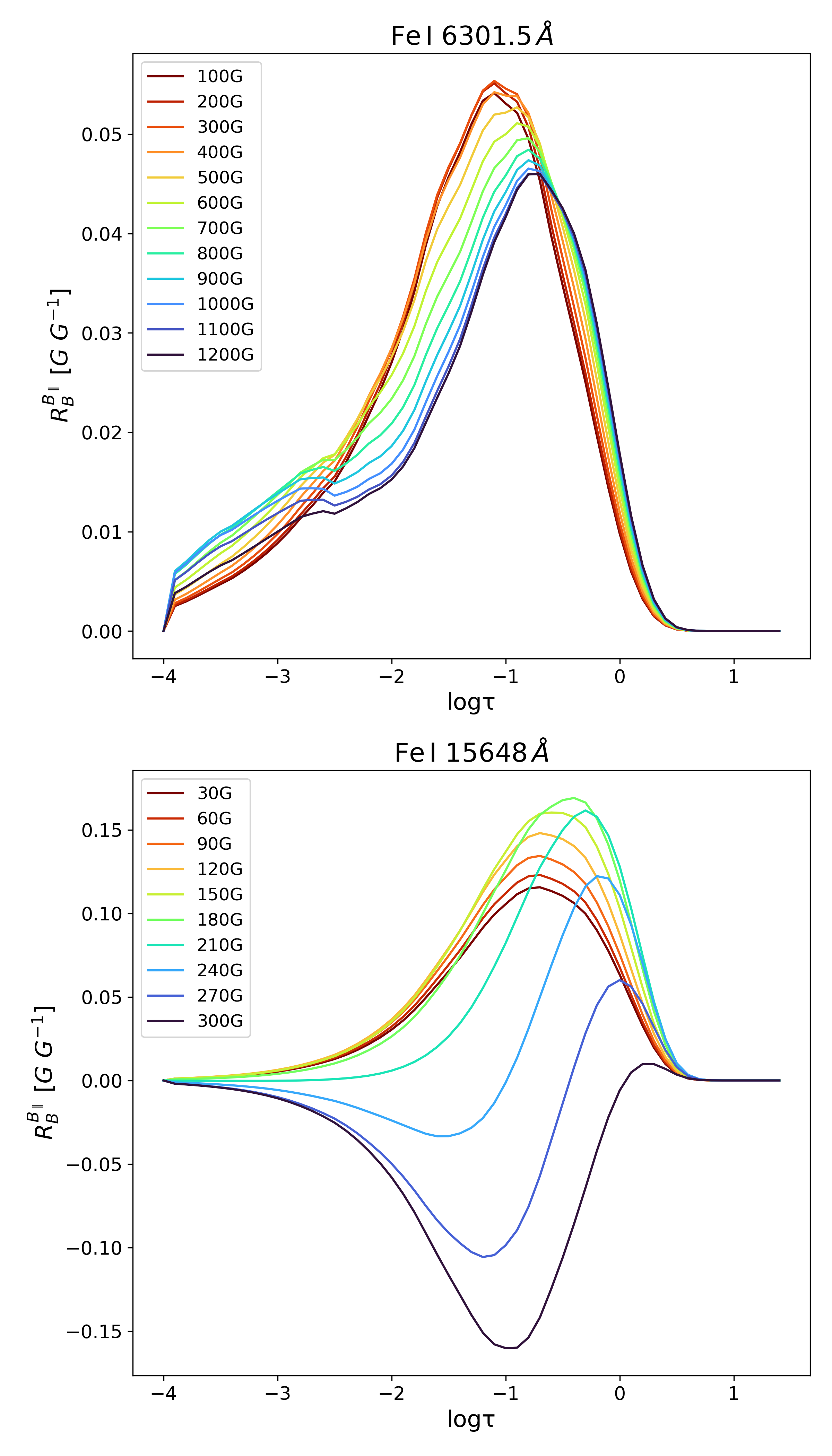}
    \caption{Generalized response functions of the longitudinal WFA ($B_{\parallel}$) as a function of $\log\tau$ for \ion{Fe}{i} 6301.5~\AA~ and 15648~\AA, for different magnetic field strengths with $\gamma=0^\circ$.}
    \label{fig3}
\end{figure}

The identification of the partial derivatives of $V$ and $I'$ with the RFs of Stokes $V$ and $I'$ to perturbations in $x_j$ leads to the GRF expression:
\begin{equation}
    \label{eq:R_BP}
    \begin{split}
    R^{B_{\parallel}}_{x_j} = -\frac{1}{C\,\overline{g}}\,\left[\frac{\sum_{i=1}^n \left( R^{V_{i}}_{x_j}\,I_i' + V_i\,R^{I'_i}_{x_j} \right)}{\sum_{i=1}^n (I_i')^2} -\right. \\ \left. \frac{2\sum_{i=1}^n V_i I'_i \sum_{i=1}^n I'_i R^{I'_i}_{x_j}}{\left[\sum_{i=1}^n (I'_i)^2\right]^2}\right],
    \end{split}
\end{equation}
where $R^{I'}_{x_j} = \partial R^I_{x_j} / \partial \lambda$.

Similarly, the GRF for the transverse field can be obtained by differentiating $B_{\perp}^2$ from Eq.~(\ref{eq:B_WFA_T}) with respect to the parameter $x_j$. This yields (neglecting higher-order terms):
\begin{equation}
    \label{eq:R_BT}
    \begin{split}
    R^{B_{\perp}^2}_{x_j} \left( \lambda_{\rm core} \right)
    = - \frac{4}{\,C^2\,\overline{G}}\,\left[\frac{\sum_{i=1}^n \Big( R^{Q_i}_{x_j}\,I_i'' + Q_i\,R^{I_i''}_{x_j} \Big)}{\sum_{i=1}^n \left(I_i''\right)^2} \right. - \\ \left.\frac{2\sum_{i=1}^n Q_i I''_i \sum_{i=1}^n I''_i R^{I''_i}_{x_j}}{\left[\sum_{i=1}^n (I''_i)^2\right]^2}\right], 
    \\
    R^{B_{\perp}^2}_{x_j} \left( \lambda_{\rm wing} \right)
    = \frac{4}{\,3C^2\,\overline{G}}\,\left[\frac{\sum_{i=1}^n \Big( R^{Q_i}_{x_j}\,I_i^* + Q_i\,R^{I_i^*}_{x_j} \Big)}{\sum_{i=1}^n \left(I_i^*\right)^2} - \right. \\ \left.\frac{2\sum_{i=1}^n Q_i I^*_i \sum_{i=1}^n I^*_i R^{I^*_i}_{x_j}}{\left[\sum_{i=1}^n (I^*_i)^2\right]^2}\right], 
    \end{split}
\end{equation}
where $I_i^*$ is defined as $I_i^* \equiv \frac{I_i'}{(\lambda - \lambda_0)}$, showing the same mathematical structure for both expressions and for Equation~(\ref{eq:R_BP}). Following the chain rule, the response to $B_{\perp}$ is:
\begin{equation}
    \label{eq:R_BT_2}
    R^{B_{\perp}}_{x_j} = \frac{1}{2B_{\perp}} R^{B_{\perp}^2}_{x_j},
\end{equation}
wherever $B_\perp \neq 0$.
In the following sections, we use these expressions to compute the GRFs numerically and assess the validity of the approximation under different physical conditions.

\subsection{Longitudinal field GRFs} \label{sec:longitudinal}

We first examine the GRF of the WFA's $B_{\parallel}$ to perturbations in the true field strength, $R^{B_\parallel}_B$. Figure~\ref{fig3} shows these GRFs as a function of the logarithmic optical depth of the continuum, $\log\tau$, for two photospheric lines ({\ion{Fe}{i} 6301.5~\AA~ and \ion{Fe}{i} 15648~\AA) and several magnetic field strengths, as obtained with the SIR code. 

Here we use $\gamma = 0^\circ$. As we already showed, this is the best-case scenario. We will further discuss the effects of inclination in Section~\ref{sec:inclination}. The GRF values at given optical depths, shown in Fig. \ref{fig3}, indicate how much the inferred $B_\parallel$ changes (in G) when the true field is perturbed by 1 G at a narrow vicinity of that depth (hence the units $\mathrm{G}\, \mathrm{G}^{-1}$). For both lines, the maximum absolute response occurs around $\log\tau \approx -1$, indicating the atmospheric layers where the spectral line is most sensitive to perturbations in the magnetic field strength through the WFA. This does not imply that the sensitivity is restricted to those layers; both lines (and, in general, all lines) are sensitive to perturbations of the physical quantities in a broad range of layers.

The maximum sensitivity of the \ion{Fe}{i} 6301.5~\AA\, line shifts toward deeper layers as $B$ increases. This means that the WFA yields different values for real atmospheres with different field-strength stratifications with depth. Also, the amplitude of the response decreases with increasing $B$. This reflects a gradual loss of sensitivity of the WFA as the magnetic field strength departs from the weak-field regime. The \ion{Fe}{i} 15648~\AA line exhibits, however, an increase in the amplitude of the RF, which reflects the overestimation of the true magnetic field also seeing in Fig. \ref{fig1}. Quite remarkably and---maybe for some---surprisingly, this line also displays negative responses for fields stronger than or equal to 200 G: this should imply a change of polarity!

Should the WFA be an ideally unbiased method---that is, one that when the measured quantity is constant ($B_\mathrm{LOS} = B_0$) yields that constant---the inferred $B_\parallel$ would verify
\begin{equation}
    \label{eq:idealmeas}
    B_\parallel = B_0 \int_0^\infty R^{B_\parallel}_B \, \text{d}\tau \equiv B_0\, \overline{R^{B_\parallel}_B},
\end{equation}
because $B_0$ is independent of depth and the integral of the GRF is not normalized, according to the theoretical derivation by \cite[][see also \citeauthor{del Toro 03} \citeyear{del Toro 03}]{Almeida 96} and the empirical verification by \cite{Westendorp Plaza 98}. Therefore, the integrated GRF quantifies a scale factor between the inferred  $B_{\parallel}$ and the true (constant) $B_0$. Hence, negative GRFs imply polarity changes.

The predicted values from Eq.~(\ref{eq:idealmeas}) are plotted in Fig.~\ref{fig1} in blue and red dashed lines. Note that above $\sim 200$ G for the visible line and above $\sim 100$ G for the IR one, WFA inferences fail to reproduce the theoretical prediction, clearly indicating that the assumptions under which the approximation might entail an unbiased measurement method break down. As a matter of fact, the predictions for the IR line imply negative polarity inferences for fields stronger than 200 G.

Figure~\ref{fig4} shows the integrated GRF, $\overline{R^{B_\parallel}_B}$, for a set of seven spectral lines as a function of $B$ for the HSRA model with $\gamma = 0$. For weak magnetic fields, the responses are close to unity, indicating that the WFA correctly retrieves the longitudinal magnetic field in this regime.  However, as the magnetic field strength increases, the integrated response progressively shifts, leading first to a overestimation followed by a systematic underestimation of the true longitudinal field. Notably, in the case of the \ion{Fe}{i} 15648 \AA\ line errors greater than 20 \% are expected for fields greater than $\sim 100$ G. For lines such as \ion{Fe}{i} 6301.5 \AA\ or \ion{Si}{i} 10827 \AA, this 20 \% error threshold occurs at much stronger field values.

\begin{figure}
    \centering
    \includegraphics[width=\hsize]{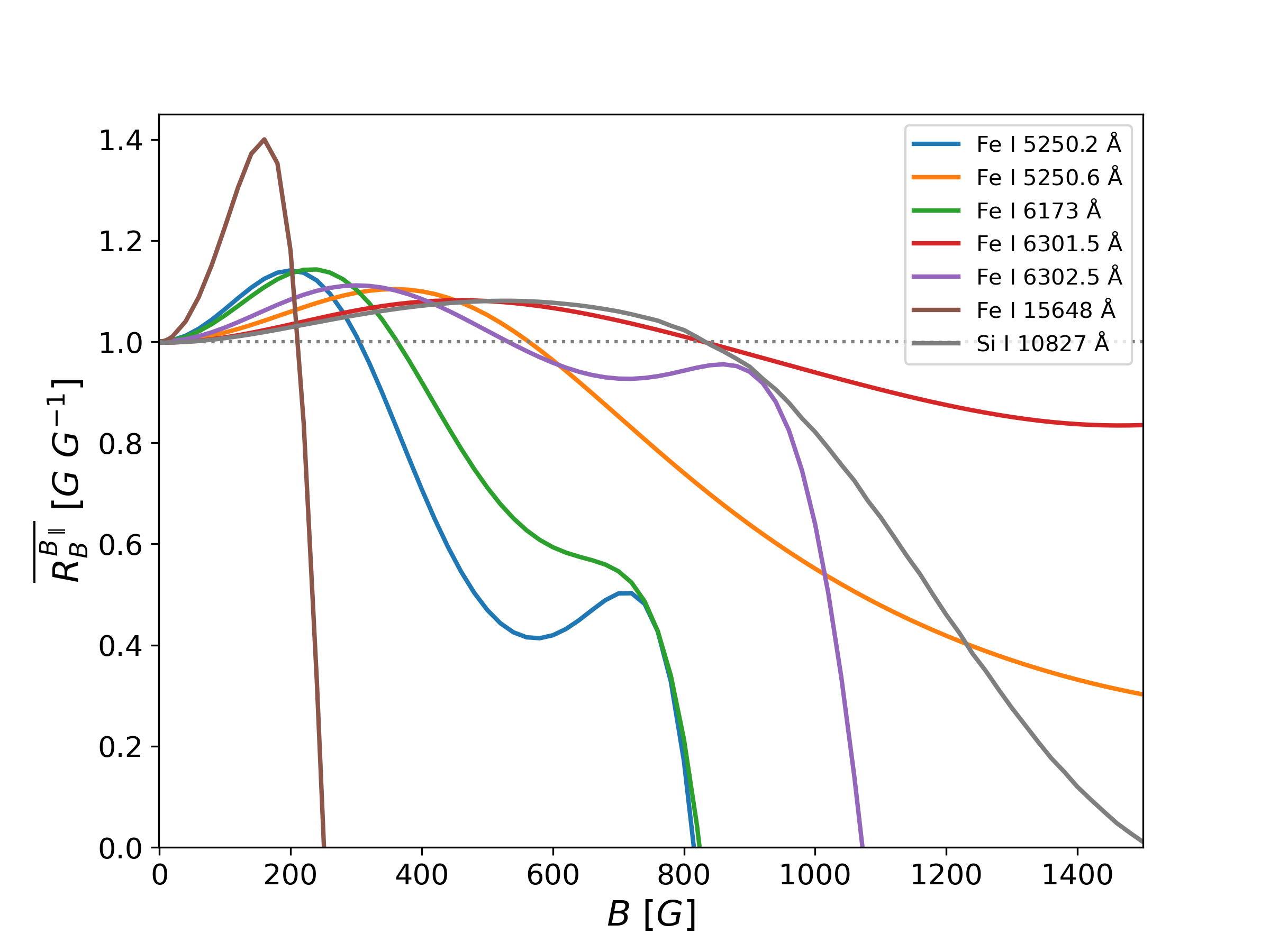}
    \caption{Integrated GRFs of $B_{\parallel}$ for seven spectral lines. }
    \label{fig4}
\end{figure}

Importantly, this behavior appears well before reaching the theoretical limit $B_{\rm sat}$ (as it should be according to \citeauthor{Landi 04} \citeyear{Landi 04}). This result shows that the practical range of validity of the WFA is significantly narrower than often assumed. As discussed in Sect.~2, this limitation arises from the breakdown of the linear relationship between $V$ and the derivative of $I$, which is one of the fundamental features of the approximation.

In the next subsection we extend this analysis by examining the sensitivity of the inferred $B_{\parallel}$ to variations in the magnetic field inclination through the corresponding GRFs.

\subsection{Inclination effects} \label{sec:inclination}

\begin{figure*}
    \centering
    \begin{subfigure}{0.49\textwidth}
        \centering
        \includegraphics[width=\textwidth]{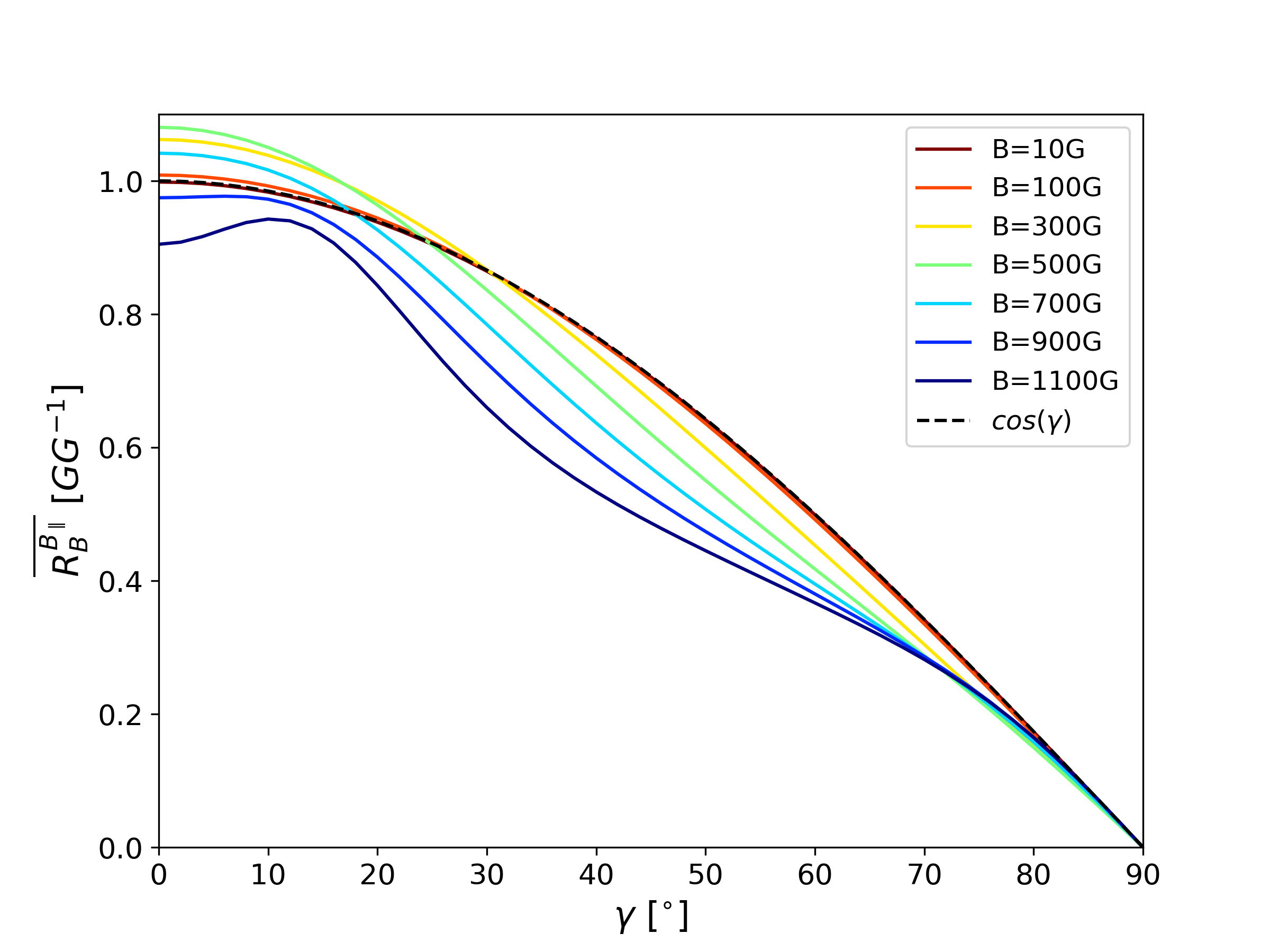}
    \end{subfigure}
    \begin{subfigure}{0.49\textwidth}
        \centering
        \includegraphics[width=\textwidth]{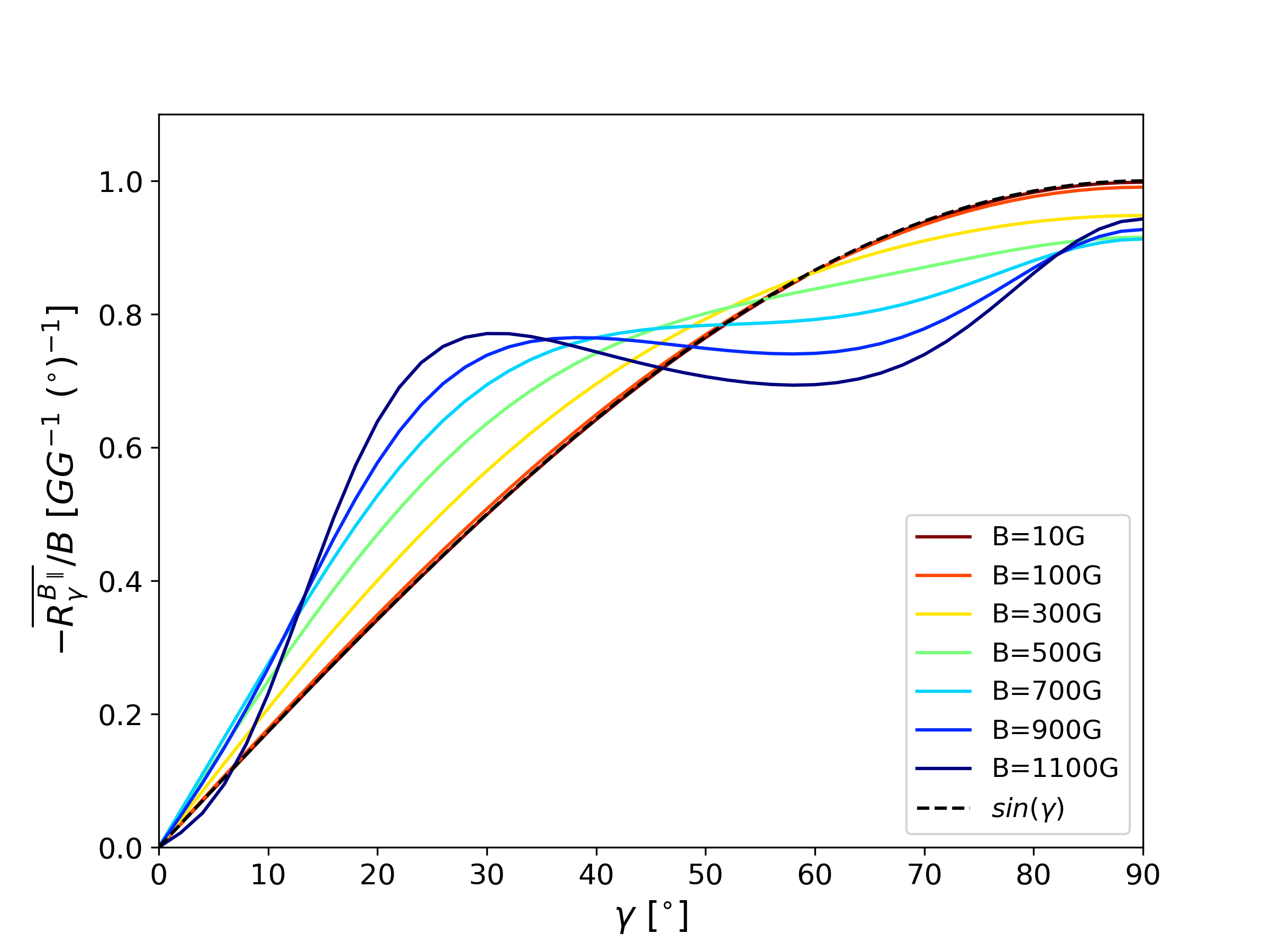}
    \end{subfigure}
    \caption{Integrated GRFs of $B_{\parallel}$ with respect to perturbations in $B$ (left) and in $\gamma$ (right), as functions of $\gamma$ and $B$.  The left plot follows $\delta B_{\parallel} \approx \cos\gamma\,\delta B$ (Eq.~\ref{eq:deltaBwf_B}), while the right follows $\delta B_{\rm WFA} \approx -B\sin\gamma\,\delta\gamma$ (Eq.~\ref{eq:deltaBwf_gamma}). Only at small $\gamma$ does the WFA accurately recover $B_{\parallel}$.}
    \label{fig5}
\end{figure*}

Contrary to what has been assumed sometimes, the WFA does depend on the magnetic field inclination as exemplified in the numerical results of Figs.~\ref{fig1} and \ref{fig2}. Although it pursues an estimation of the longitudinal field, its applicability cannot be checked by varying only $B$ while fixing $\gamma = 0$. The inclination of the magnetic field significantly affects the results. In fact, imagine that the approximation is perfect: $B_{\parallel}=B\cos\gamma$. In such a case, modifications in $B_{\parallel}$ can come from perturbations in both $B$ and $\gamma$: 
\begin{equation}
    \label{eq:deltaBwf}
    \delta B_{\parallel} = \delta(B\cos\gamma) = \cos\gamma\,\delta B - B\sin\gamma\,\delta\gamma.
\end{equation}
If we set $\delta\gamma=0$ and only vary $B$, then
\begin{equation}
    \label{eq:deltaBwf_B}
    \delta B_{\parallel} = \cos\gamma\,\delta B,
\end{equation}
the response of the inferred field to $B$ would be scaled by $\cos\gamma$. On the other hand, if only $\gamma$ changes ($\delta B=0$), then
\begin{equation}
    \label{eq:deltaBwf_gamma}
    \delta B_{\parallel} = -B\sin\gamma\,\delta\gamma,
\end{equation}
the response to inclination perturbations is proportional to $-B\sin\gamma$. 

Equations (\ref{eq:deltaBwf_B}) and (\ref{eq:deltaBwf_gamma}) imply that for small inclinations ($\gamma\approx0$), the inferred $B_{\parallel}$ is essentially insensitive to small changes in $\gamma$. However, for larger $\gamma$ or large-enough $B$, the $\sin\gamma$ term becomes significant. This trend is seen in Fig.~\ref{fig1}: even moderate inclinations (30°--45°) lead to large discrepancies in the inferred $B_{\parallel}$ at relatively weak magnetic fields.

Figure~\ref{fig5} shows the GRFs with respect to $B$ and $\gamma$ for different magnetic field strengths as a function of the inclination. The left panel shows $\overline{R^{B_\parallel}_B}$. The right panel displays the sensitivity to small perturbations in $\gamma$, $-\overline{R^{B_\parallel}_\gamma}/B$. As expected, when the field strength is small enough, the GRFs approach the ideal behaviors from Eq.~(\ref{eq:deltaBwf_B}) and (\ref{eq:deltaBwf_gamma}). However, as the magnetic field strength increases, the response progressively departs from the ideal dependence.

In real observations, since the true inclination is generally unknown, neglecting the inclination effect can lead to unreliable $B_\parallel$ inferences except in the very weak-field regime.  

\subsection{Transverse field GRFs} \label{sec:transverse}

We now extend the previous analysis to the transverse component of the magnetic field. In the WFA, the transverse field is inferred from the linear polarization signals, which involve higher-order terms in the weak-field expansion. 

As discussed earlier, it is necessary to distinguish between the line-core and line-wing regimes when applying the two transverse-field expressions. For the core regime, the second expression in Eq.~(\ref{eq:WFA_B}) is strictly valid only at $\lambda=\lambda_0$. However, to obtain a more robust estimate, we use a least-squares fit to a narrow wavelength range of $\pm10\,\mathrm{m\AA}$ around $\lambda_0$. In contrast, the wing regime is less sharply defined. As beautifully explained in \citet{Landi 04}, Eq.~(\ref{eq:WFA_Q_wings}) is valid under the conditions $|\lambda-\lambda_0| \gg \Delta\lambda_D$ and $|\lambda-\lambda_0| \gg \lambda_0^2 \Gamma^\prime/c$, where $\Gamma^\prime$ is the damping parameter of the line; therefore, we adopted twice the line FWHM ($\sim \Delta\lambda_\mathrm{D}$) as a safe boundary for this regime. Although a factor 2 may not seem very accurate to comply with the requirement, note that the adopted wavelength samples do already belong to the far wings of the line.

In practice, this requirement or a similar one may induce that the estimation of the transverse-field component can barely be done with one or two wavelengths if the spectral sampling is not very fine. If the requirement is not fulfilled, then the observer must be acquainted of the intrinsic inaccuracy of such a procedure. It is also worth noting that, in the far-wing regime ($|\lambda-\lambda_0| \gg \Delta\lambda_D$), the linear polarization signals are intrinsically weak. As a consequence, instrumental noise is expected to have a significant impact on the inferred field strengths in real observations, an effect that has not been considered in the present work.

\begin{figure}[h]
    \centering
    \includegraphics[width=\hsize]{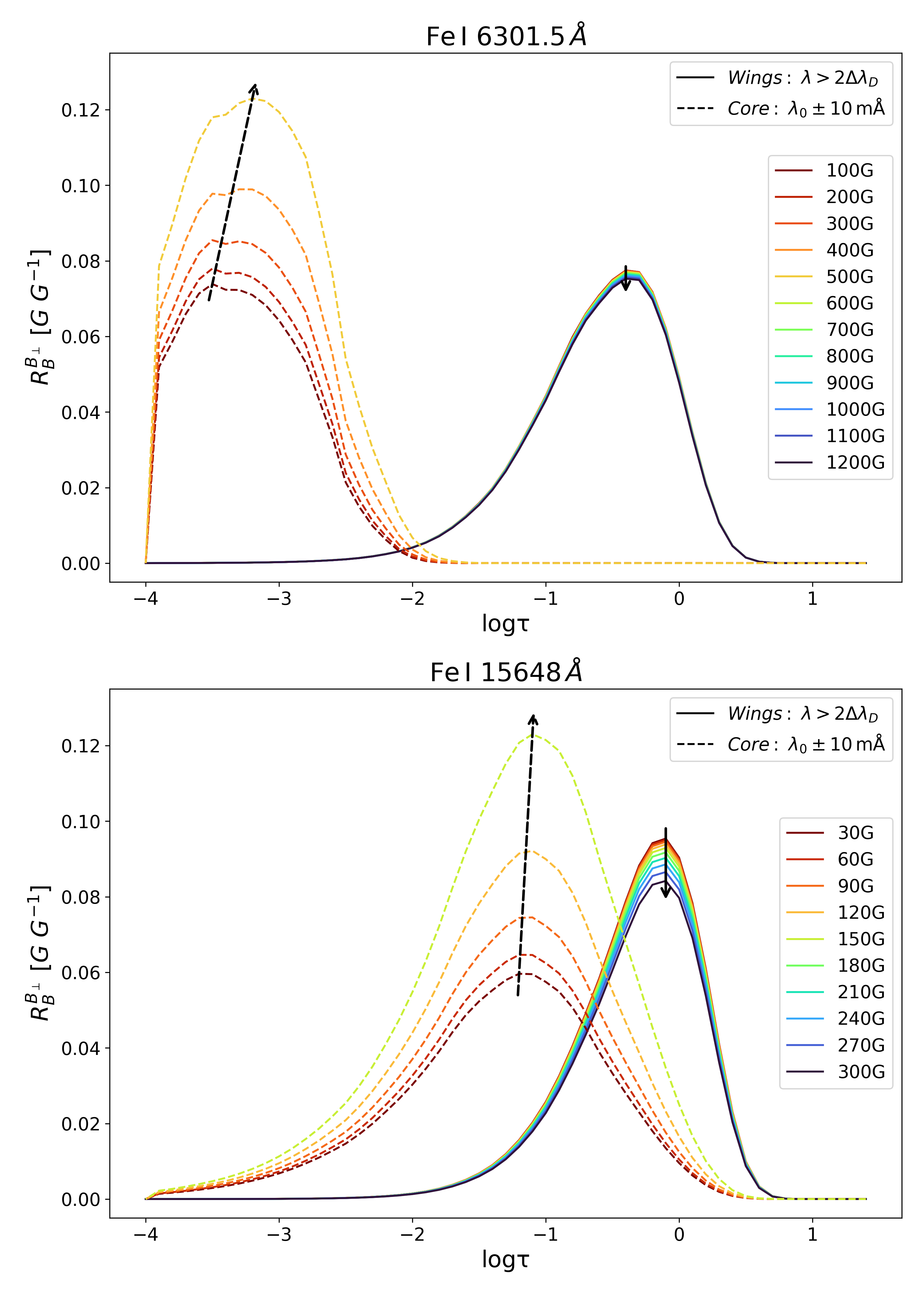}
    \caption{GRFs of $B_{\perp}$ with respect to $B$ for the \ion{Fe}{i} lines at 6301.5~\AA~ and 15648~\AA\ with $\gamma = 90^\circ$, computed in the core (dashed) and wings (solid) regimes. Arrows show the trend of the maximum when increasing $B$.}
    \label{fig6}
\end{figure}

Figure~\ref{fig6} shows the corresponding GRFs computed using the line-core and line-wing expressions. For simplicity, we adopted a magnetic-field inclination of $\gamma = 90^\circ$. The response obtained from the wings (solid lines) exhibits a maximum near $\log\tau = 0$---even deeper than the longitudinal field---and a decreasing amplitude as $B_\mathrm{T}$ increases, as in the longitudinal case. In contrast, the response associated with the line core displays the opposite trend: it increases with an increasing $B_\mathrm{T}$. Furthermore, the maximum response of the core and wing regimes occurs at different atmospheric heights. This effect is particularly pronounced for the \ion{Fe}{i} line at 6301.5~\AA, whose core response peaks around $\log\tau \sim -3$ and remains significant at $\log\tau < -4$, and beyond the range covered by the atmospheric model used in this work. These differences in height tell by themselves what has been repeatedly claimed \citep{Ruiz Cobo 94, del Toro 03}: spectral lines are not formed at given heights. Within a single line, different wavelengths are sensitive to the various solar physical quantities at different heights. And the differences can even apply to different quantities: these GRFs for the transverse component peak at different optical depths than those for the longitudinal magnetic field. This warning may sound too academic when assuming, as in our case, that $\vec{B}$ is constant with $\tau$. However, the real Sun does not know about our simplifications. Real-world inferences can be proven to be averages through the whole atmosphere of the real quantity ($B_\mathrm{LOS}$, $B_\mathrm{T}$, $v_\mathrm{LOS}$, etc.) weighted by the corresponding response function \citep{Almeida 96, del Toro 03}.

As in the longitudinal case, the relevant diagnostic is obtained by integrating the GRFs over logarithmic optical depth because our $B_\mathrm{T}$ is assumed to be constant throughout the atmosphere. For clarity, the depth-dependent GRFs are shown in Fig.~\ref{fig6} only for two representative spectral lines, whereas the integrated responses were computed for a set of seven lines. 

In Fig.~\ref{fig7} we show the quantity $\overline{R^{B_{\perp}}_B}$, which directly corresponds to the theoretical ratio between the inferred and true transverse magnetic fields, $B_\perp/B_\mathrm{T}$, as a function of the actual field strength. ($B = B_\mathrm{T}$ because the inclination is kept to $\gamma= 90^\circ$.) The integrated responses in the top panel indicate that the transverse WFA expression is able to recover the true transverse field up to quite high magnetic field strengths, even for the two \ion{Fe}{i} lines at 5250.2 \AA\ and 6173 \AA, those with the higher effective Landé factors of our sample in the visible. However, as expected, the IR line fails relatively soon in providing a reliable inference.

\begin{figure}
    \centering
    \includegraphics[width=\hsize]{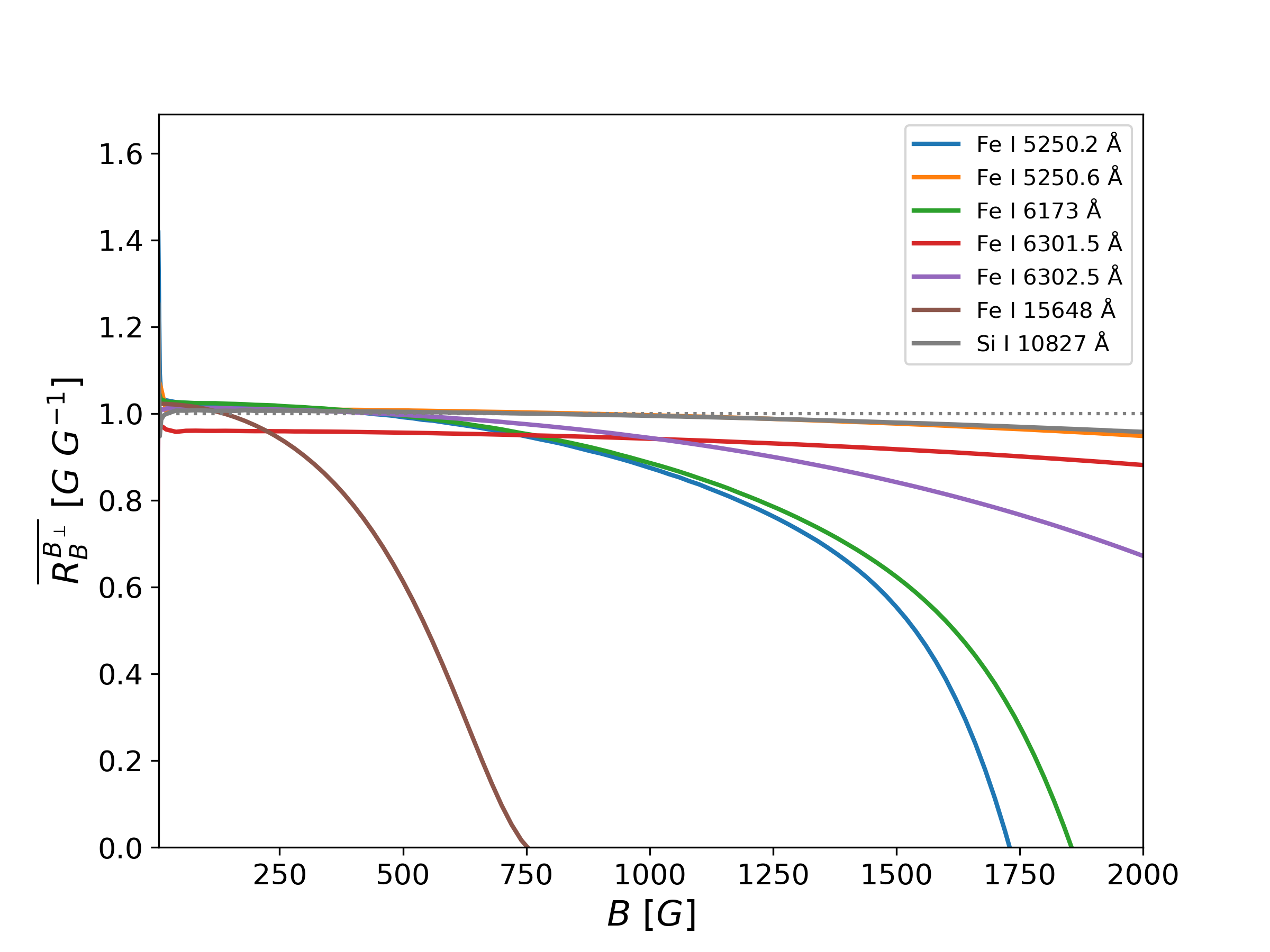}
    \includegraphics[width=\hsize]{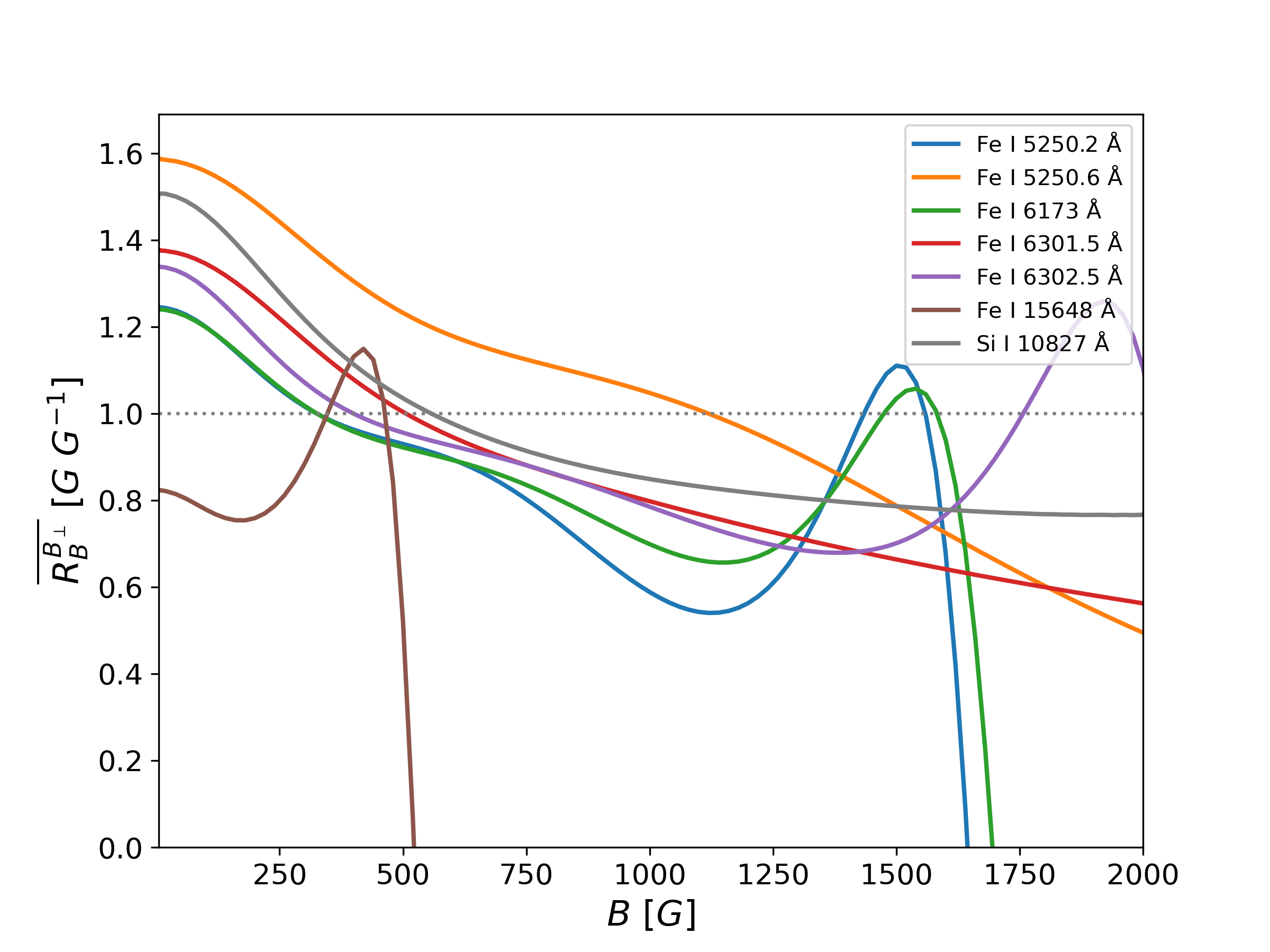}
    \includegraphics[width=\hsize]{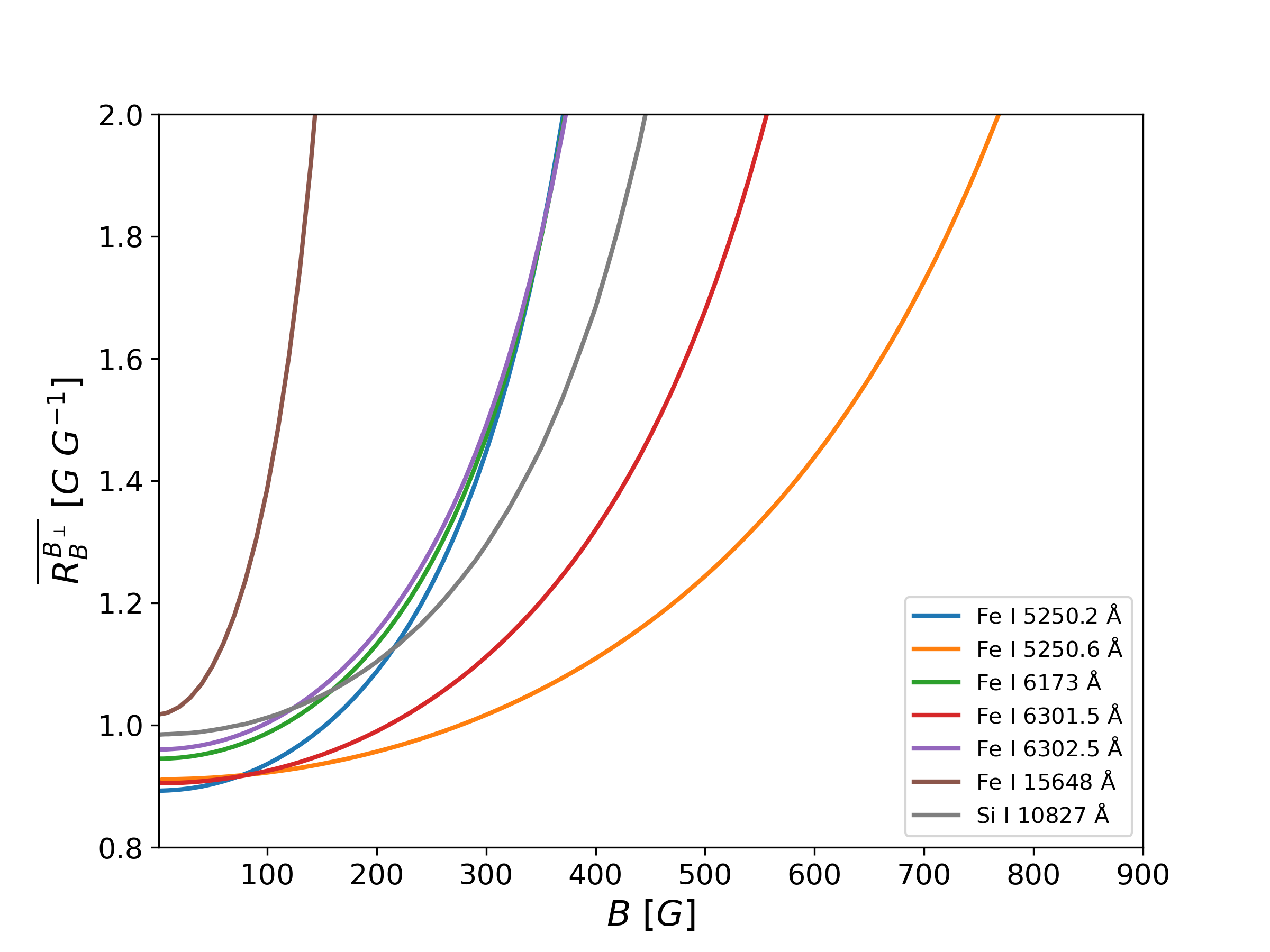}
    \caption{Integrated GRFs of the transverse WFA for seven spectral lines. Top panel: using the wing expression for $|\lambda-\lambda_0| > 2\, \mathrm{FWHM}$. Middle panel: using the wing expression up to the the inflection point of the lines. Bottom panel: using the core expression for $\lambda_0 \pm 10 \,\mathrm{m}\AA$. The curves represent the theoretical ratio between the transverse magnetic field inferred by the WFA and the true transverse magnetic field (which is the true magnetic field strength as $\gamma=90^\circ$). }
    \label{fig7}
\end{figure}

While the previous analysis adopted a wing definition based on the formal validity conditions of Eq.~(\ref{eq:WFA_Q_wings}), namely $|\lambda-\lambda_0| > 2~\mathrm{FWHM}$ ($\sim \Delta\lambda_\mathrm{D}$), alternative criteria have also been used in the literature \citep[e.g.,][]{Centeno 18}. In this approach, the boundary between the line core and the wings is defined by the inflection point of the intensity profile. Although this criterion was meant to somehow separating the photospheric and chromospheric information conveyed by the \ion{Ca}{ii} line at 8542 \AA---which seems a fairly reasonable assumption,\footnote{Indeed, it is a good assumption for the longitudinal case where no specific additional requirement exists. See Section~\ref{sec:cromospheric}.} it might provide an operationally natural separation between core and wing regimes for all lines while retaining larger polarization signals for the inference. This would make measurements potentially less sensitive to instrumental noise. 

Despite the inflection-point criterion may not fulfill the conditions $|\lambda-\lambda_0| \gg \Delta\lambda_D$ and $|\lambda-\lambda_0| \gg \lambda_0^2 \Gamma^\prime/c$, and knowing from the analysis of the longitudinal case that large deviations from linearity occur precisely near the inflection points, we thought it was worth the trial and explore the potential sensitivity to the transverse field component of the WFA's $B_\perp$ with this separation.

The middle panel of Fig.~\ref{fig7} shows the integrated responses obtained when the line wings are defined using the intensity inflection points. The inferred transverse magnetic fields exhibit large departures from the true values, even at very weak fields with all the lines in our sample. Although this criterion provides stronger polarization signals and may appear more attractive from an observational perspective, the deterioration in performance confirms that the wing expression is being applied outside its optimal range of validity. This confirms our result in Sect.~2, that using different wavelength samples returns completely different values of the searched-for quantity. 

Lastly, the bottom panel shows the integrated responses obtained using the core expression applied within the line core range ($\lambda_0 \pm 10\, \mathrm{m}\AA$). The core wavelength samples rapidly lead to an overestimation, ultimately failing to recover the true transverse magnetic field indeed at quite moderate strengths, even for fully horizontal fields: errors larger than 20 \% occur for $B \geq 400$ or 500 G for the more accurate lines. The tendency to increase $B_\perp$ with $B_\mathrm{T}$ could have been foreseen from Fig.~\ref{fig6}. The origin of this behavior will become clearer after Sect.~\ref{sec:revisited}, as it arises from the dependence of Stokes $I$ on the magnetic field strength and the consequent splitting, which are neglected by the regular implementations of the WFA. 

These theoretical results have been confirmed with some numerical experiments. The reader can be sure that they highlight the importance of selecting wavelength intervals that are consistent with the assumptions of the approximation, even at the expense of reduced signal amplitudes. This trade-off between maximizing the validity of the WFA in the line wings and preserving sufficient polarimetric signal has long been recognized in the literature. \citet{Jefferies 91} already pointed out that, although the approximation is expected to perform better in the far-wings, the weaker polarization signals often limit its practical application.

\begin{figure*}
    \centering
    \begin{subfigure}{0.49\textwidth}
        \centering
        \includegraphics[width=\textwidth]{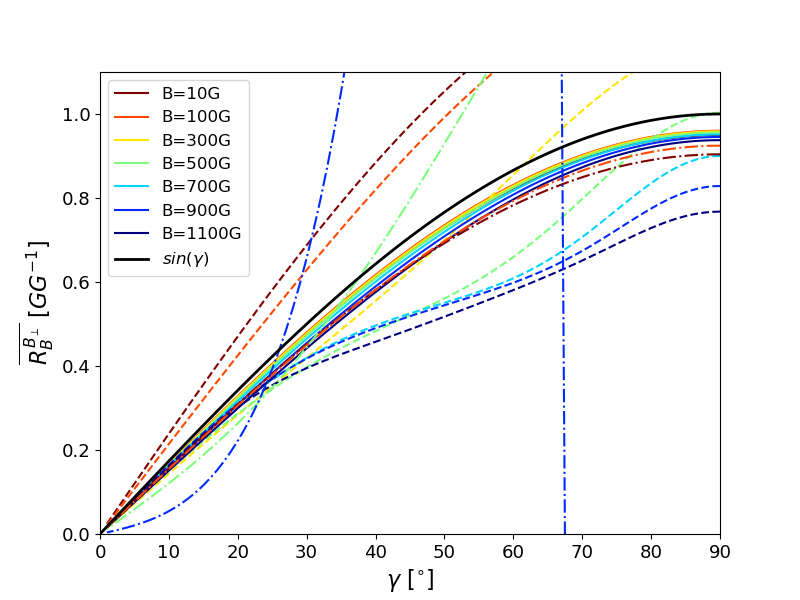}
    \end{subfigure}
    \begin{subfigure}{0.49\textwidth}
        \centering
        \includegraphics[width=\textwidth]{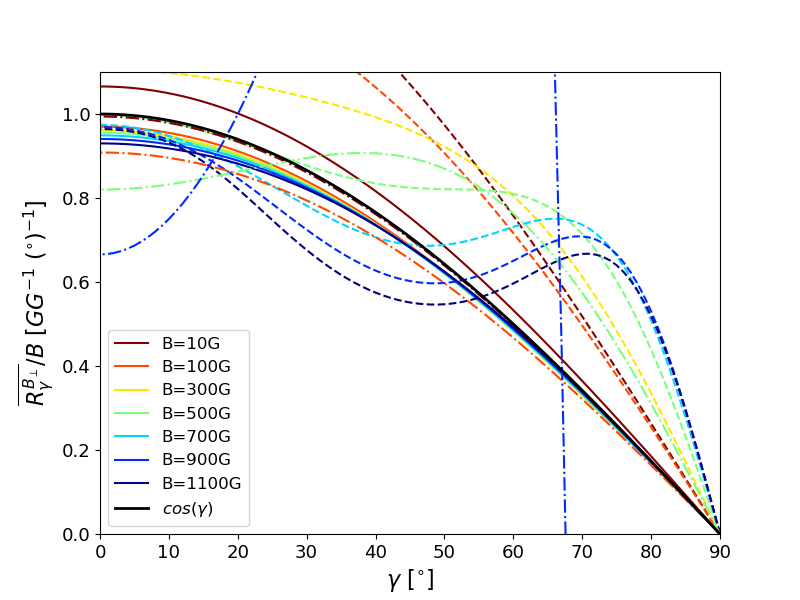}
    \end{subfigure}
    \caption{Integrated GRFs of $B_{\perp}$ with respect to perturbations in $B$ (left) and in $\gamma$ (right), as functions of $\gamma$ for different field strengths. The different line styles correspond to the different wavelength regimes: far-wings (solid), wings (dashed), core (dashed-dotted). The black solid line in the left panel follows $\delta B_{\perp} = \sin\gamma\,\delta B$, while that in the right panel follows $\delta B_{\perp} = -B\cos\gamma\,\delta\gamma$.}
    \label{fig8}
\end{figure*}

\subsection{Inclination effects}
\label{sec:inclitran}

Similarly to the longitudinal case, the WFA for the transverse magnetic field also depends on the magnetic field inclination. Since the transverse component is given by $B_\perp=B\sin\gamma$, changes in $B_\perp$ can arise from variations in both the magnetic field strength and inclination: 
\begin{equation}
    \label{eq:deltaBwf_T}
    \delta B_{\perp} = \delta(B\sin\gamma) = \sin\gamma\,\delta B + B\cos\gamma\,\delta\gamma
\end{equation}
We can set $\delta\gamma=0$ or $\delta B=0$ to separately study these effects. The results are shown in Fig.~\ref{fig8}) for the far-wing (solid lines), up to the inflection point (dashed lines), and the core (dashed-dotted lines). 

The responses to perturbations in the magnetic field strength (left panel) decrease as $\gamma \rightarrow 0^\circ$. Using the far-wings (solid), the numerical GRFs closely follow the expected $\sin\gamma$ dependence. In contrast, when the inflection point is adopted as the wings threshold (dashed), the response departs significantly from the ideal behavior, even for very weak and horizontal fields. The core wavelength samples (dashed-dotted) reproduce the $\sin\gamma$ dependence only for sufficiently weak fields, but then rapidly deviates from it as the magnetic field strength increases.

The response to perturbations in the inclination (right panel) exhibits a similar behavior as regards the wavelength ranges. The expected $\cos\gamma$ is almost well recovered by the far-wing range regardless of the magnetic field strength (up to 1000 G at least), while the core range only reproduces it for sufficiently weak fields. In contrast, the range from the inflection point outwards shows much bigger departures from the ideal dependence for even weak fields.

Overall, these results reinforce the conclusions obtained for the longitudinal component, while adding a new layer of complexity with the separation in wavelength regimes. As stated at the end of Sect.~\ref{sec:formulation},  using different wavelength samples will return completely different magnetic fields.

\subsection{Chromospheric lines} \label{sec:cromospheric}

Chromospheric lines form over a broad range of atmospheric heights, and therefore their GRFs exhibit a more complex behavior. Following the analysis presented in the previous section, we investigate these lines using the three wavelength regimes (core, wings, and far-wings), with the aim of disentangling the magnetic field contributions originating from different heights in the solar atmosphere.

The chromospheric lines considered here, \ion{Ca}{ii} 8542~\AA\ and \ion{Mg}{i} $b_2$ 5173~\AA\, were synthesized using the DeSIRe \citep{Ruiz Cobo 22} radiative transfer code with the FALC \citep{Fontenla 1993} model atmosphere. Synthetic Stokes profiles were computed for magnetic field strengths ranging from 10 to 1000 G, considering both vertical ($\gamma = 0^\circ$) and horizontal ($\gamma = 90^\circ$) magnetic fields, and azimuth $\varphi=0^\circ$.

Both \ion{Ca}{ii} 8542~\AA\ and \ion{Mg}{i} $b_2$ 5173~\AA\ exhibit a much clearer separation between the core and the wing regimes. We applied the WFA independently to the line core ($\lambda_0\pm50$ m\AA\ for \ion{Ca}{ii} 8542~\AA\ and $\lambda_0\pm30$ m\AA\ for \ion{Mg}{i} $b_2$), the wings (starting at the inflection points), and the far-wings (beyond only one Doppler width of the line, $\Delta\lambda_D$, not two).

The resulting GRFs, shown in Fig.~\ref{fig9}, reveal a clear height separation between the line-core and wing regimes for both the longitudinal ($B_\parallel$) and transverse ($B_\perp$) magnetic field calculations. In both spectral lines, the core is primarily sensitive to magnetic fields around $\log\tau \approx -5$, corresponding to chromospheric layers, whereas the wings and far-wings probe deeper atmospheric regions. As a result, these wavelength regimes can also be used to estimate the photospheric magnetic field.

In the longitudinal case, the wing and far-wing GRFs look very similar, probing almost the same heights. For the transverse scenario, however, the far-wing GRFs appear significantly noisier. This behavior is purely numerical and stands from the extremely weak linear polarization signals present in the far wings. Although some differences exist between the wing and far-wing responses, as expected from the dependence of the WFA on the selected wavelength range, both regimes remain predominantly sensitive to atmospheric layers at $\log\tau > -3$.

A particularly interesting result is that the photospheric magnetic field inferred from the wings of these chromospheric lines can be comparable to, or even more reliable than, that obtained from dedicated photospheric lines. This behavior is likely a consequence of the much broader profiles of the chromospheric lines, which provide a larger wavelength range over which the WFA remains applicable while still having polarization signals in the far-wings regime. 

\begin{figure*}
    \centering
    \includegraphics[width=\hsize]{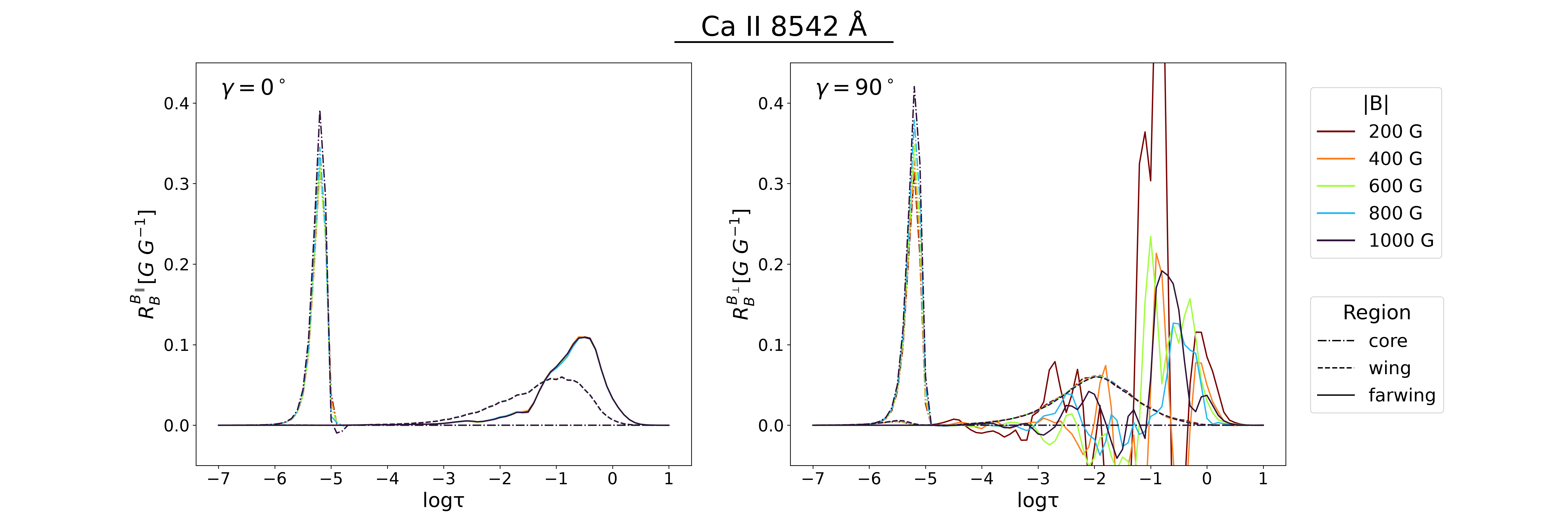}
    \includegraphics[width=\hsize]{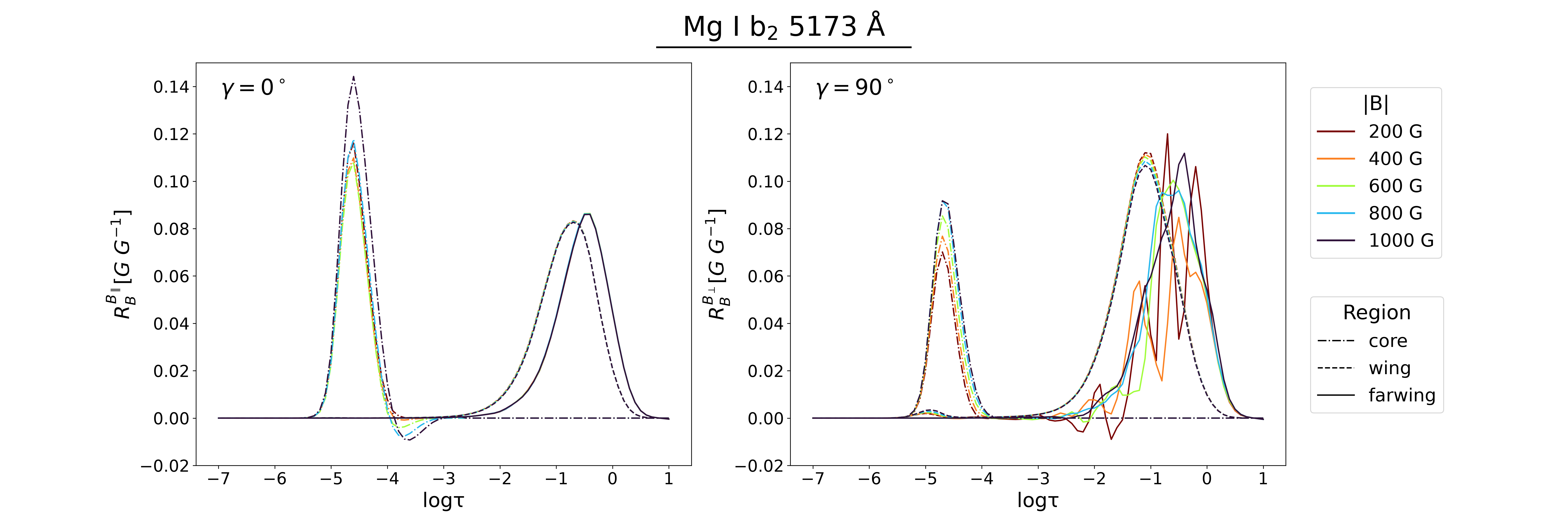}
    \caption{GRFs of $B_{\parallel}$ (left panels) to $B$ with $\gamma=0^\circ$ and GRFs of $B_{\perp}$ (right panels) to $B$ with $\gamma=90^\circ$ for two representative chromospheric lines. The different line styles correspond to the different wavelength regimes: far-wings (solid), wings (dashed), core (dashed-dotted). }
    \label{fig9}
\end{figure*}

Finally, in the next section we revisit the original formulation of the WFA to understand how these limitations arise from the assumptions underlying the approximation itself.

\section{The WFA revisited} \label{sec:revisited}

The elegant formulation of the Weak Field Approximation by \citet{Landi 73} is based on a perturbative solution of the radiative transfer equation (RTE) for polarized light in the presence of a magnetic field. In this formalism, the introduction of an artificial parameter $\lambda$ (called $\alpha$ in \citeauthor{Landi 04} \citeyear{Landi 04})---which multiplies the vector magnetic field $\vec{B}$ and is finally set to one at the end---allows the Stokes parameters to be expanded in series of powers of $v_B$. Substituting these series into the RTE, systems of equations can be obtained for different orders of approximation. 

Specifically, if the macroscopic LOS velocity and the vector magnetic field are assumed constant throughout the atmosphere, Eqs.~(\ref{eq:WFA_formal}) are deduced. An important remark is in order, namely, that they do not correspond to the same order of approximation. This is an important feature for the correctness of their practical application to observations.

To be more precise, both equations are formally correct but \textbf{cannot be consistently used in practice simultaneously}. The equation for Stokes $V$ is only valid up to first order of approximation and that for Stokes $Q$ is valid up to second order, but the often forgotten expansion of Stokes $I$ readily tells us that the $I$ appearing in both equations either is not the same in the two equations or does not coincide with the observed Stokes $I$. This is clear in the first formulation \citep{Landi 73} but the unfortunate notation---yet legitimate and elegant in any other respect---specified in Eq.~(9.73) of \cite{Landi 04} may induce to misleading interpretation in practical applications. Stokes $I$ in our Eqs.~(\ref{eq:WFA_formal}) must be $I_0$, that is, the zeroth-order approximation of Stokes $I$ or, in plain words, the equivalent intensity of an atmosphere equal to that under consideration but with no magnetic field.\footnote{As beautifully remarked by \cite{Landi 04}, having an atmosphere with the same source function in the absence of a magnetic field is not trivial for lines formed out of local thermodynamic equilibrium, which is the case for many chromospheric lines.}

\begin{table}[h]
\caption{Stokes parameter expansion terms in various orders of approximation}
\label{tab:approx}
\begin{tabular}{ccccc}
\hline \\[-8pt]
Order & Stokes $I$ & Stokes $Q$ & Stokes $U$ & Stokes $V$ \\
\hline
0 & $I_0$ & 0 & 0 & 0 \\
1 & $I_0$ & 0 & 0 & $V_1$ \\
2 & $I_0 + I_2$ & $Q_2$ & $U_2$ & $V_1$ \\
3 & $I_0 + I_2$ & $Q_2$ & $U_2$ & $V_1 + V_3$ \\
$\vdots$ \\
\hline
\end{tabular}
\end{table}

To help clarify this issue, let us summarize the different orders of approximation in Table~\ref{tab:approx}. In the table, we follow the same convention as in the two aforementioned works: $I_0, I_1, I_2, \ldots$, $Q_0, Q_1, Q_2, \ldots$, $U_0, U_1, U_2, \ldots$, and $V_0, V_1, V_2, \ldots$, while the observable Stokes parameters follow from the sum of all the terms up to a given order. As clearly seen in the table, at first order of approximation, both the observable $V$ and $I$ are those needed for the first of Eqs.~(\ref{eq:WFA_formal}), but if we accept up to the second order, none of the equations can be applied as the observable Stokes $I = I_0 + I_2$. As already claimed, e.g., by \cite{del Toro 16}, as soon as linear polarization is observed, the so-called magnetograph formula is not strictly valid and can only be used if the field is really weak. Only then one can expect that $I_2$ is small enough as for writing $I \simeq I_0$. Note that this is clearly specified by \cite{Landi 04}: when $v_B \rightarrow 0$. Indeed, the linear polarization can be expected of the order of $I_2$. Only when Stokes $Q$ and $U$ are very small, can $I_2$ be very small.

Therefore, the breakdown of the approximation is not primarily related to the amplitude of the polarization signals, but rather to the modification of the intensity profile itself as the field strength increases, which alters the shape and derivative of the Stokes $I$ profile. As $B$ (or $\gamma$) grows, the observed intensity progressively departs from $I_0$, and the proportionality between the observed polarization signals and the derivative of the intensity is no longer strictly satisfied.

There are still some considerations in order regarding the second-order approximation. As commented before in Sect.~\ref{sec:transverse}, Eqs.~(\ref{eq:WFA_Q_wings}) and the second of Eqs.~(\ref{eq:WFA_B}) are valid under the same set of assumptions only if the former is applied to the far wings of the line, where $|\lambda-\lambda_0| \gg \Delta\lambda_D$ and $|\lambda-\lambda_0| \gg \lambda_0^2 \Gamma^\prime/c$. This is because the extra assumption needed for using a further expression \citep[Eq. 9.84 in][]{Landi 04} for the linear polarization, which might be valid at any wavelength except for the core of the line, is hard to fulfill except in the far wings: it requires that both $\Delta\lambda_\mathrm{D}$ and $\Gamma^\prime$ are constant with depth. Notably, there, in the far wings of the line, both the smallness of the linear polarization and the validity of the approximation equation hold.\footnote{\citet{Centeno 18} wrongly states that the second of Eqs.~(\ref{eq:WFA_B}) is valid for any $\lambda$ except for the core of the line.} Quite remarkably, there is a second-order approximation consequence that can be used regardless of the intensity. Equation (9.81) in \cite{Landi 04}, $\tan 2\varphi = U(\lambda)/Q(\lambda)$ in our notation, is valid to second order and can be applied in practice because it only relies on measured quantities.

These considerations show that the limitations discussed in the previous sections are not merely practical issues of the approximation, but are already implicit in its original formulation. The perturbative derivation assumes that magnetic effects introduce only small corrections to the non-magnetic intensity profile and that the linear relations between the Stokes parameters and the magnetic field remain valid. When these conditions are no longer satisfied, the proportionality relations that form the basis of the WFA progressively break down. The behavior revealed by the response functions therefore reflects the intrinsic limits of the approximation rather than shortcomings of its implementation.

\section{Conclusions} \label{sec:conclusions}

In this work we have revisited the validity of the Weak Field Approximation (WFA) and analyzed its limitations using synthetic spectra and response-function diagnostics. Our results show that the linear relation between the Stokes $V$ profile and the derivative of the intensity progressively breaks down as the magnetic field departs from the weak-field regime. This result explains the loss of linearity between $V$ and $I'$ discussed in Sect.~2 and illustrated in Fig.~\ref{fig2}, and therefore the uncertainties in the calculation of the magnetic field. Even when the WFA fit still appears to reproduce the real magnetic field reasonably well, the inferred magnetic field may no longer represent the actual field strength.

In addition, the WFA relations implicitly assume that the inferred longitudinal field corresponds to the projection of the magnetic field vector along the line of sight. The quantity retrieved through the WFA is therefore $B_{\parallel}=B\cos\gamma$, which depends not only on the magnetic field strength but also on its inclination. As shown in Sect.~3, this introduces an additional source of sensitivity to variations in $\gamma$, so that changes in inclination can significantly affect the inferred $B_{\parallel}$ even when the magnetic field strength remains unchanged.

Moreover, the transverse WFA faces even more severe limitations. Its application relies on additional assumptions which restrict its use to specific spectral regions, either at  line center or sufficiently far in the wings. As a consequence, its practical implementation is inherently more constrained and sensitive to deviations from this ideal scenario. Most importantly, as shown in Sect.~4, the transverse WFA introduces an intrinsic bias that arises directly from the original mathematical formulation of the approximation. The transverse WFA should therefore be applied with particular caution and only after carefully evaluating whether the assumptions underlying the approximation are fulfilled for the specific spectral line and magnetic configuration under consideration.

In summary, our results show that the effective range of validity of the WFA is more limited than is often assumed. Although the approximation can still provide useful estimates in the weak-field regime, departures from the underlying assumptions can introduce significant biases in the inferred magnetic field. The analysis presented here highlights the importance of carefully assessing the validity of the WFA for each spectral line and magnetic configuration before interpreting the retrieved magnetic fields.

A final warning is in order: all calculations in this work have been done for a magnetic azimuth $\varphi = 0$. Should $\vec{B}$ be constant with depth, no further problem is expected from this assumption. In real applications, where this condition is not expected to hold, the non-linear relationship $L^2 = Q^2 + U^2$ introduces an extra layer of difficulties, due to the non-linear character of its definition.

\begin{acknowledgements}
This work have been funded by the Spanish AEI/MCIN/10.13039/501100011033/ projects PID2021-125325OB-C51, CNS2023-144723, and PID2024-156066OB-C5, and from “Center of Excellence Severo Ochoa” awards to IAA-CSIC (CEX2021-001131-S), all co-funded by European REDEF funds, “A way of making Europe”. A.L.S.T. acknowledges funding from the Consejer\'ia de Transformaci\'on Econ\'omica, Industria, Conocimiento y Universidades of the Junta de Andaluc\'ia through grant POSTDOC-21-00832.
\end{acknowledgements}


\begin{thebibliography}{}

\bibitem[Afonso Delgado et al.(2023)]{Alfonso 23} Afonso Delgado, D., del Pino Alem{\'a}n, T., \& Trujillo Bueno, J.\ 2023, \apj, 942, 2, 60. doi:10.3847/1538-4357/aca669
\bibitem[Afonso Delgado et al.(2025)]{Afonso 25} Afonso Delgado, D., del Pino Alem{\'a}n, T., Trujillo Bueno, J., et al.\ 2025, \apj, 991, 2, 164. doi:10.3847/1538-4357/adfcd4
\bibitem[Asensio Ramos (2011)]{Asensio Ramos 11} Asensio Ramos, A. 2011, \apj, 731, 27. doi: 10.1088/0004-637X/731/1/27
\bibitem[Asensio Ramos et al.(2008)]{Asensio Ramos 08} Asensio Ramos, A., Trujillo Bueno, J., \& Landi Degl'Innocenti, E.\ 2008, \apj, 683, 542. doi:10.1086/589433
\bibitem[Bellot Rubio \& Orozco Su{\'a}rez(2019)]{Bellot y Orozco 19} Bellot Rubio, L. \& Orozco Su{\'a}rez, D.\ 2019, Living Reviews in Solar Physics, Quiet Sun magnetic fields: an observational view, 16, 1, 1. doi:10.1007/s41116-018-0017-1
\bibitem[Centeno(2018)]{Centeno 18} Centeno, R.\ 2018, \apj, 866, 89. doi:10.3847/1538-4357/aae087
\bibitem[Centeno et al.(2010)]{Centeno 10} Centeno, R., Trujillo Bueno, J., \& Asensio Ramos, A.\ 2010, \apj, 708, 1579. doi:10.1088/0004-637X/708/2/1579
\bibitem[Cobos Carrascosa (2015)]{Cobos 2015} Cobos Carrascosa, J.~P.\ 2015, High-Performance Scientific Computing on FPGA aboard the Solar Orbiter PHI Instrument, PhD thesis, University of Granada, Spain. https://digital.csic.es/handle/10261/137976
\bibitem[Del Toro Iniesta(2003)]{del Toro 03} Del Toro Iniesta, J.~C.\ 2003, Introduction to spectropolarimetry. Cambridge University Press: Cambridge. 
\bibitem[Del Toro Iniesta \& Ruiz Cobo(2016)]{del Toro 16} Del Toro Iniesta, J.~C. \& Ruiz Cobo, B.\ 2016, Living Reviews in Solar Physics, 13, 1, 4. doi:10.1007/s41116-016-0005-2
\bibitem[D{\'\i}az Baso et al.(2025)]{Diaz Baso 25} D{\'\i}az Baso, C.~J., Asensio Ramos, A., de la Cruz Rodr{\'\i}guez, J., et al.\ 2025, \aap, 693, A170. doi:10.1051/0004-6361/202452172
\bibitem[Dom{\'\i}nguez Cerde{\~n}a et al.(2003)]{Dominguez 03} Dom{\'\i}nguez Cerde{\~n}a, I., Kneer, F., \& S{\'a}nchez Almeida, J.\ 2003, \apjl, Quiet-Sun Magnetic Fields at High Spatial Resolution, 582, 1, L55. doi:10.1086/346199
\bibitem[Ferrente et al.(2023)]{Ferrente 23} Ferrente, F., Zuccarello, F., Guglielmino, S.~L., et al.\ 2023, \apj, 954, 185. doi:10.3847/1538-4357/ace9d1
\bibitem[Fontenla et al.(1993)]{Fontenla 1993} Fontenla, J.~M., Avrett, E.~H., \& Loeser, R.\ 1993, \apj, 406, 319. doi:10.1086/172443
\bibitem[Gingerich et al.(1971)]{Gingerich 71} Gingerich, O., Noyes, R.~W., Kalkofen, W., et al.\ 1971, \solphys, 18, 347. doi:10.1007/BF00149057
\bibitem[Jefferies \& Mickey(1991)]{Jefferies 91} Jefferies, J.~T. \& Mickey, D.~L.\ 1991, \apj, 372, 694. doi:10.1086/170012
\bibitem[Kleint(2017)]{Kleint 17} Kleint, L.\ 2017, \apj, 834, 1, 26. doi:10.3847/1538-4357/834/1/26
\bibitem[Landi Degl'Innocenti \& Landi Degl'Innocenti(1973)]{Landi 73} Landi Degl'Innocenti, E. \& Landi Degl'Innocenti, M.\ 1973, \solphys, 31, 299. doi:10.1007/BF00152807
\bibitem[Landi Degl'Innocenti \& Landi Degl'Innocenti(1977)]{Landi 77} Landi Degl'Innocenti, E. \& Landi Degl'Innocenti, M.\ 1977, \aap, 56, 111
\bibitem[Landi Degl'Innocenti \& Landolfi(2004)]{Landi 04} Landi Degl'Innocenti, E. \& Landolfi, M.\ 2004, Polarization in Spectral Lines. By Egidio Landi Degl'innocenti and Marco Landolfi, University of Fierenze, Firenze, Italy; Arcetri Observatory, Firenze, Italy. ASTROPHYSICS AND SPACE LIBERARY Volume 307 Kluwer Academic Publishers, Dordrecht. doi:10.1007/978-1-4020-2415-3
\bibitem[Mart{\'\i}nez Gonz{\'a}lez \& Bellot Rubio(2009)]{Martinez Gonzalez 09} Mart{\'\i}nez Gonz{\'a}lez, M.~J. \& Bellot Rubio, L.~R.\ 2009, \apj, 700, 1391. doi:10.1088/0004-637X/700/2/1391.
\bibitem[Martínez Pillet et al. (1990)]{Pillet 90} Martinez Pillet, V., García López, R.J., Del Toro Iniesta, J.C., Vázquez, M., Beckman, J.E., \& Char, S.\ 1990, \apjl, 361, L81.
\bibitem[Orozco Su{\'a}rez et al.(2015)]{Orozco Suarez 15} Orozco Su{\'a}rez, D., Asensio Ramos, A., \& Trujillo Bueno, J.\ 2015, \apjl, 803, L18. doi:10.1088/2041-8205/803/2/L18
\bibitem[Pietarila et al.(2007)]{Pietarila 07} Pietarila, A., Socas-Navarro, H., \& Bogdan, T.\ 2007, \apj, 670, 885. doi:10.1086/521271
\bibitem[Ruiz Cobo(1998)]{Ruiz Cobo 98} Ruiz Cobo, B.\ 1998, \apss, 263, 331. doi:10.1023/A:1002191418914
\bibitem[Ruiz Cobo et al.(2022)]{Ruiz Cobo 22} Ruiz Cobo, B., Quintero Noda, C., Gafeira, R., et al.\ 2022, \aap, 660, A37. 
doi:10.1051/0004-6361/202140877
\bibitem[Ruiz Cobo \& Del Toro Iniesta(1992)]{Ruiz Cobo 92} Ruiz Cobo, B. \& Del Toro Iniesta, J.~C.\ 1992, \apj, Inversion of Stokes Profiles, 398, 375. doi:10.1086/171862
\bibitem[Ruiz Cobo \& Del Toro Iniesta(1994)]{Ruiz Cobo 94} Ruiz Cobo, B. \& Del Toro Iniesta, J.~C.\ 1994, \aap, 283, 129.
\bibitem[Sánchez Almeida (1992)]{Almeida 92} Sánchez Almeida, J.\ 1992, \solphys, 137, 1.
\bibitem[Sánchez Almeida, Ruiz Cobo, \& Del Toro Iniesta (1996)]{Almeida 96} Sánchez Almeida, J., Ruiz Cobo, B., \& Del Toro Iniesta, J.C.\ 1996, \aap, 314, 295.
\bibitem[Semel (1967)]{Semel 1967} Semel, M.\ 1967, Annales d'Astrophysique, 30:513-513.
\bibitem[Shchukina \& Trujillo Bueno (2019)]{Shchukina 19} Shchukina, N.G. \& Trujillo Bueno, J.\ 2019, \aap 628, A47.
\bibitem[Siu-Tapia et al.(2020)]{Siu Tapia 20} Siu-Tapia, A.~L., Bellot Rubio, L.~R., Orozco Su{\'a}rez, D., et al.\ 2020, \aap, 642, A128. doi:10.1051/0004-6361/202038370
\bibitem[Siu-Tapia et al.(2025a)]{Siu Tapia 25a} Siu-Tapia, A.~L., Bellot Rubio, L.~R., Orozco Su{\'a}rez, D., et al.\ 2025, \aap, 696, A105. doi:10.1051/0004-6361/202453232
\bibitem[Siu-Tapia et al.(2025b)]{Siu Tapia 25b} Siu-Tapia, A.~L., Bellot Rubio, L.~R., Orozco Su{\'a}rez, D., et al.\ 2025, \aap, 696, A106. doi:10.1051/0004-6361/202453232
\bibitem[Stein(2012)]{Stein 12} Stein, R.~F.\ 2012, Living Reviews in Solar Physics, 9, 1, 4. doi:10.12942/lrsp-2012-4
\bibitem[Takeda \& UeNo(2017)]{Takeda 17} Takeda, Y. \& UeNo, S.\ 2017, \pasj, 69, 46. doi:10.1093/pasj/psx022
\bibitem[Vukadinovi{\'c} et al.(2022)]{Vukadinovic 22} Vukadinovi{\'c}, D., Mili{\'c}, I., \& Atanackovi{\'c}, O.\ 2022, \aap, 664, A182. doi:10.1051/0004-6361/202142015
\bibitem[Westendorp Plaza et al.(1998)]{Westendorp Plaza 98} Westendorp Plaza, C., Del Toro Iniesta, J.~C., Ruiz Cobo, B., et al.\ 1998, \apj, 494, 453. doi:10.1086/305192

\end{thebibliography}
\end{document}